\documentclass[aps,pre, amssymb,longbibliography]{revtex4-2}
\usepackage{amsmath}  
\usepackage{graphicx}  

\begin{document}
\title{Scattering Signatures of Invasion Percolation}

\author{Jean-Christian Angl\`es d'Auriac}
\affiliation{
  Universit\'e Grenoble Alpes, Institut N\'eel, F-38042 Grenoble, France\\
  CNRS, Institut N\'eel, F-38042 Grenoble, France
}

\author{Pierre-Etienne Wolf}
\email{pierre-etienne.wolf@neel.cnrs.fr}
\affiliation{
  Universit\'e Grenoble Alpes, Institut N\'eel, F-38042 Grenoble, France\\
  CNRS, Institut N\'eel, F-38042 Grenoble, France
}

\date{\today}

\begin{abstract}
Motivated by recent experiments, we investigate the scattering properties of percolation clusters
generated by numerical simulations on a three dimensional cubic lattice. 
Individual clusters of given size are shown to present a fractal structure up to a scale of order their extent, even far away from the percolation threshold  $p_c$. The influence of inter-cluster correlations on the structure factor of assemblies of clusters selected by an invasion phenomenon is studied in detail. For invasion from bulk germs, we show that the scattering properties are determined by three length scales, the correlation length $\xi$, the average distance between germs $d_g$, and the spatial scale probed by scattering, set by the inverse of the scattering wavevector $Q$. At small scales, we find that the fractal structure of individual clusters is retained,  the structure factor decaying as $Q^{-d_f}$. At large scales, the structure factor tends to a limit, set by the smaller of  $\xi$ and $d_g$, both below and above $p_c$.  We propose approximate expressions reproducing the simulated structure factor for arbitrary $\xi$, $d_g$, and $Q$, and illustrate how they can be used to avoid to resort to costly numerical simulations.  For invasion from surfaces, we find that, at $p_c$, the structure factor  behaves as $Q^{-d_f}$ at all $Q$,\textit{ i.e.} the fractal structure is retained at arbitrarily large scales. Results away from $p_c$ are compared to the case of bulk germs. Our results can be applied to discuss light or neutrons scattering experiments on percolating systems. This is illustrated in the context of evaporation from porous materials. 
\end{abstract}

\maketitle
\section{Introduction}
\label{sec:intro}
Take a lattice. Pick up sites at random and group these sites into clusters 
of connected sites (so-called Bernoulli clusters). When the fraction $p$ of chosen sites approaches 
a critical value $p_c$, the size of the largest
cluster diverges. This phenomenon is known as site percolation. Similarly,  bond percolation occurs when bonds rather than sites,  are considered.
Percolation, either site or bond, is ubiquitous in science.
 In mathematics, it has  been widely studied by
mathematicians as a example of conformal invariance and a
realization of a Schramm-Loewner evolution. In geology, it is used to describe the transport of fluids (oil in particular) in porous rocks. In statistical physics, percolation provides the
simplest model of a phase transition. Examples of applications are numerous and include some problems in magnetism, electrical transport in disordered alloys or granular superconductors, gelation of polymers, growth models, transport in complex networks, \ldots . We refer the reader to Ref. \citenum{Saberi2015a} for references. 

\medskip
Percolation has spectacular consequences in terms of transport properties. A classical example of bond percolation is the random resistor network when the fraction of conductive bonds is increased. Below percolation, no continuous path exists between two sides of the sample, and the sample is insulating. At percolation and above, a connected cluster spans the sample, and the sample conducts.  However, this conduction threshold is not the only feature of percolation. In particular, at the percolation point, the percolating cluster has a fractal structure. The  fractal dimension $d_f$  depends on the lattice dimensionnality, but not on its precise geometry. Smaller (finite) clusters have the same fractal structure, but only up to the scale of their extent. 

\medskip 
It could thus be expected that light or neutrons scattering, which probe the spatial correlations between chosen sites, can detect percolation. However, a key point makes the problem more subtle that could be assumed at a first glance. Radiation scattering is sensitive to the spatial distribution of sites, but \textit{not} to the connectivity of clusters. Hence, when two neighboring clusters merge into a larger one because one of the sites of their common boundary becomes picked, the scattered field barely changes. As a result, coherent light scattering (we specialize to this case from now on) cannot be expected to be as sensitive as transport to percolation.

\medskip
In fact, if \textit{all} clusters are considered  (the so-called Bernoulli problem), light scattering is totally unsensitive to percolation! Indeed, since sites are picked at random, the scattered field corresponds to that of a random distribution of scatterers. Although, for a given cluster, the field is enhanced with respect to incoherent scattering due to intra-cluster correlations, destructive interferences between clusters due to inter-cluster correlations exactly compensate this enhancement. 

\medskip 
This insensitivity of static light scattering to percolation has been previously pointed out in the context of the sol-gel transition \cite{DeGennes1979,  Cazabat1980, Martin1987b}. Due to this effect, probing the fractal clusters generated by gelation requires either to use dynamical light scattering, or to dilute the clusters so as suppress the inter-cluster correlations  \cite{Martin1988}.  In other percolation related problems, an alternative to dilution is the selection of clusters. If, among all possible Bernoulli clusters, only a fraction is effectively selected due to some physical process, then coherent effects can be restored.

\medskip
Evaporation of a fluid from disordered, 3D connected, porous materials provides examples of such a cluster selection.
These materials can often be represented as assemblies of connected cylindrical pores 
of random radius $R$ \cite{Mason1983a, Guyer1996a, Bonnet2019b}.
Starting from a situation where all pores are filled with a liquid, 
decreasing the fluid pressure will trigger evaporation. 
  As confirmed by experiments in controlled geometries \cite{Wallacher2004a, Doebele2020a}, a given pore can empty by two different mechanisms \cite{Mason1983a, Parlar1987a, Parlar1989a, Machin1999a, Ravikovitch2002a, Morishige2003a}.
   The first one consists in the displacement of a free liquid-vapor interface through the pore, which occurs provided that
   i) the pressure $P$ is smaller
   than a characteristic equilibrium pressure $P_{\rm{eq}}(R)$ decreasing with the pore radius $R$ and ii) at least one
    of its neighbors is empty so as to provide a liquid-gas interface.
The second is the thermally activated nucleation (cavitation) of a bubble 
within the pore surrounded by neighbors
filled with liquid. Cavitation occurs if the pressure becomes smaller than the cavitation pressure  $P_{\rm{cav}}(R)\leq P_{\rm{eq}}(R)$.
Both mechanisms give rise to a percolation process, the parameter $p (P)$ of percolation being the fraction of pores satisfying condition i),\textit{ i.e.} such that  $P_{\rm{eq}}(R)\geq P$.
Indeed, grouping these pores into clusters of connected sites, condition ii) implies that the pores which are effectively empty at pressure $P$ are all those belonging
  to clusters containing at least one pore emptied by cavitation ($P \leq P_{\rm{cav}}(R)$), or contacting the external free surfaces
    of the sample (in contact with the vapor). The two latter conditions respectively correspond to so-called invasion percolation from bulk germs (the cavitated pores) or from surface germs (the surface sites).
In both cases, the selection of clusters creates spatial correlations between empty pores, which should be detectable through a scattering experiment. 

\begin{figure}
\includegraphics[scale=0.3]{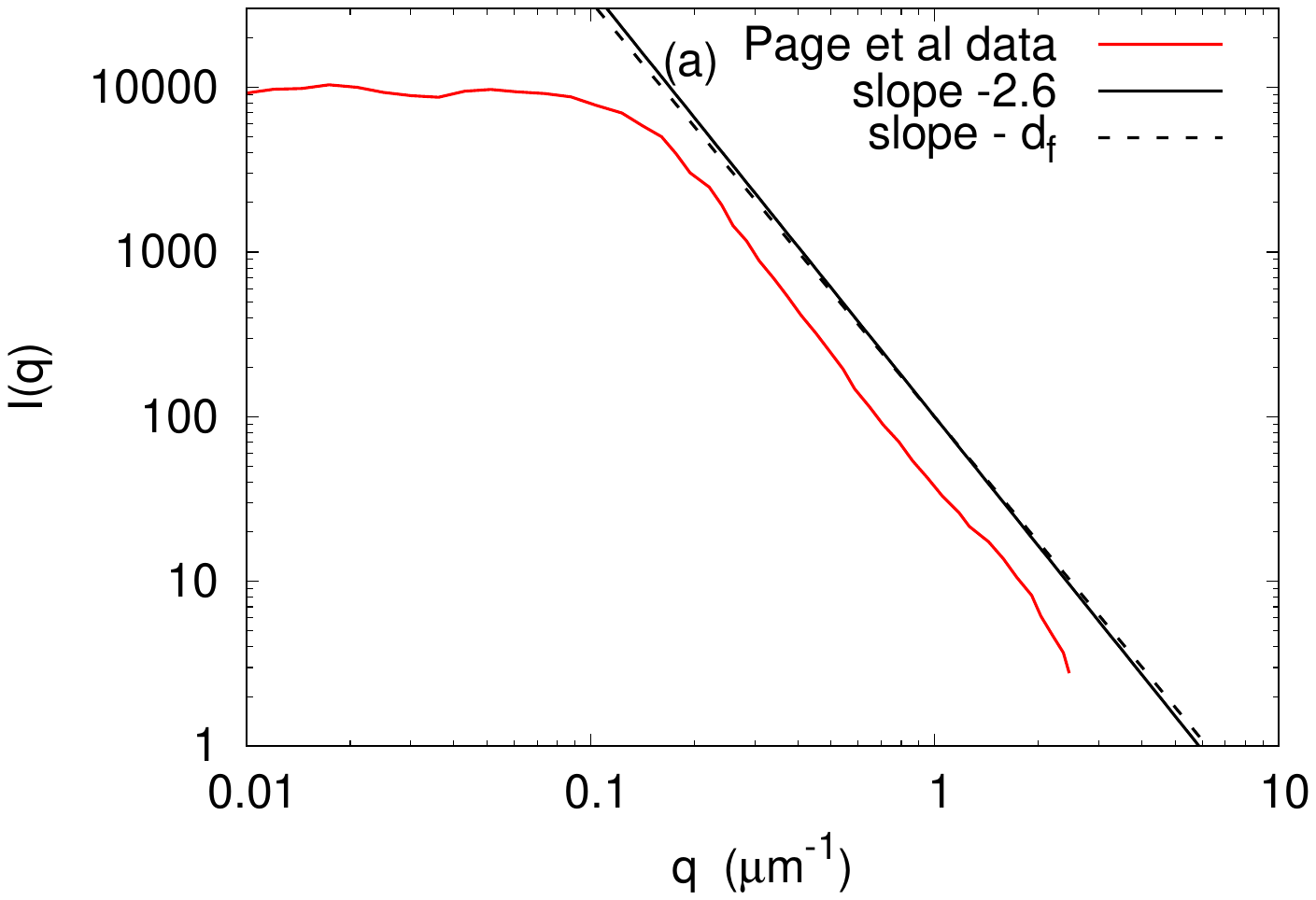} 
\includegraphics[scale=0.3]{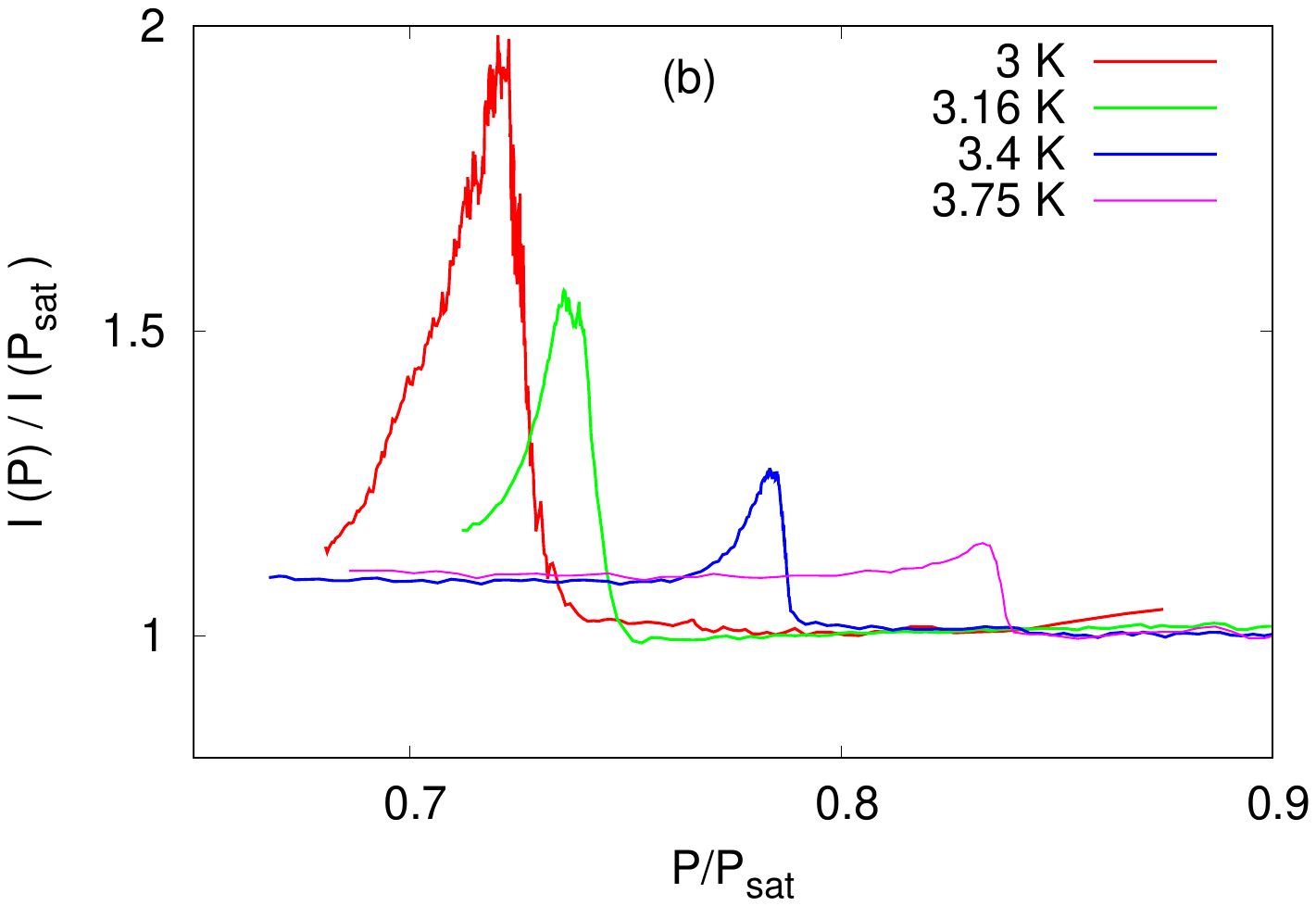} 
  \caption{ Two examples of light scattering signatures of percolation. 
 (a)  Evaporation of hexane from Vycor: power-law dependence of the scattered intensity $I(q)$ with the transfer wavevector $q$ at the precise pressure where evaporation begins (adapted from Ref.\citenum{Page1995a}). (b)  Pressure dependence of the scattered intensity $I(P)$ at 
an angle of 45$^{\circ}$ during evaporation of helium from Vycor at different temperatures (adapted from Ref.\citenum{Bonnet2013a}). The peak amplitude decreases with increasing temperature.} 
 \label{fig:Experiments}
 \end{figure}

\medskip
As illustrated by Fig.~\ref{fig:Experiments}, two previous studies using Vycor, a porous glass, as a prototypical disordered porous material, confirm this expectation. In the first one, small angle light scattering measurements\cite{Page1995a} performed during evaporation of hexane at room temperature revealed, at some well defined pressure,  a power-law dependence of the scattered intensity with the transfer wavevector $q$, with an exponent $\approx$ 2.6, close to $d_f=2.523$, the accepted fractal dimension for 3D percolation \cite{Lorenz1998,Deng2005}. This finding has been interpreted as direct evidence for evaporation being controlled by invasion percolation from the sample  surfaces. More recently, measurements performed during evaporation of liquid helium from Vycor at cryogenic temperatures have shown that the light scattered at finite angle also peaks around a temperature dependent pressure, the peak intensity decreasing with increasing temperature \cite{Bonnet2013a}. Based on direct numerical simulations of the evaporation process for the physical parameters of the problem \cite{Bonnet2019b}, this behavior has been interpreted as a signature of evaporation being controlled by invasion percolation from cavitated bulk germs.
 
 \medskip
The strength of these interpretations is however limited by the lack of a full theoretical description of the scattering signal associated with percolation, depending on the cluster selection process. Indeed, while the scaling theory of percolation gives some information on the structure of individual clusters close to the percolation threshold, it does not tell anything about assemblies of clusters, where intercorrelation effects are potentially important.  The goal of this paper is to bridge this gap by using numerical simulations to predict the scattering signatures of percolation, depending on the selection process at play.  

 \medskip
Specifically, when only the largest cluster is considered, the structure factor is expected to scale as $q^{-d_f}$ at large $q$, and  $\xi^{d_f}$, where $\xi$ is the correlation length for percolation, at small $q$. The associated small angle scattering thus peaks at $p_c$. Does this still hold when many clusters are selected, as is the case for invasion from surfaces, or from bulk germs ? And does the $q^{-d_f}$ dependence at large $q$ remain valid in the invasion regime ? Answering these questions is mandatory to be able to probe the existence of percolation by using scattering measurements.
 
\medskip
On that aim, we performed numerical simulations of percolation on a 3D cubic lattice.
In these simulations, we generate clusters at different values of $p$ and compute
 the resulting structure factor when different selection rules of clusters are applied.
   
 \medskip
 When only clusters with the same number of sites are selected,
 we essentially probe the structure factor of individual clusters. We thus determine the scattering properties of individual Bernoulli clusters, and find that their fractal structure only weakly depends
 to the proximity to the percolation threshold. 
 
 \medskip
 When the selection rule corresponds to invasion for bulk germs, we identify two different regimes. First, a dilute regime, corresponding to a density of germs small enough for the clusters to be well separated. Second, a concentrated regime corresponding to the opposite case. In the dilute limit, we show that
 the structure factor is accurately described by adding the structure factors for different cluster sizes
 weighted by the proper cluster size distribution. Because this distribution differs from the all-clusters distribution,
 the decay of the structure factor at large wavevectors is only controlled by $d_f$,
 in contrast to a result previously obtained in the context of gelation \cite{Martin_PRA1985}.
In the concentrated limit, the structure factor is depressed from the above behavior due to inter-cluster correlations, and
 we give approximate expressions 
 describing this evolution, based on the three length scales characterizing
the scattering problem, namely the correlation length, the distance between germs from which the clusters are selected, and the spatial scale probed by the transfer wavevector.  Such expressions will be useful to experimentalists using scattering to probe the percolation nature of physical problems. 

 \medskip
 Finally, we compute the scattering signal when the selection rule corresponds to invasion from surfaces, and compare it to the case of invasion from bulk germs. This allows us to discuss whether scattering experiments can discriminate these two processes.

\medskip
This paper is organized as follows. The simulation scheme and the calculation of the structure factor are discussed in section~\ref{sec:Simulation}. Section~\ref{sec:oneCluster} deals with the structure factors for different individual clusters:
 first, the largest one, and, second, clusters of given size~$s$.
  The case of percolation invasion from bulk germs is discussed in section~\ref{sec:BulkGerms}. Sections ~\ref{sec:Dilute} and \ref{sec:Concentrated} describe the results below $p_c$
   for the dilute and concentrated regime, respectively, while  
section ~\ref{sec:Abovepc} discusses the
regime above $p_c$. Section~\ref{sec:Unified} gives expressions describing the structure factor in the previous  regimes, which  we use in section~\ref{sec:Discussion} to discuss the main features of the scattering signal for invasion from bulk germs.
 Section~\ref{sec:Evaporation}  illustrates how these expressions can be used in practical cases,
  taking as an exemple the problem of evaporation from a random porous material.
  Finally, section~\ref{sec:SurfaceGerms} presents our results for invasion percolation from surfaces and compares them to the case of invasion from bulk germs.


\section{Simulation scheme}
\label{sec:Simulation}

\subsection{Generation of Bernoulli clusters}
\label{sec:Generation}
The simulations were performed at $d=3$ dimensions for a cubic lattice of linear size $L$=1024. 
In a first step, we generate Bernoulli clusters by picking sites at random with a probability $p$, where $p$ belongs to a set of 23 values distributed in an interval of  $\pm$30\% around $p_c$ (see Appendix~A for the list of $p$ values).
To this aim, we choose a random permutation of the $V=L^3$
sites, compute  for each $p$ the closest integer $k$ of $pV$ and retain
the $k$ first sites of the permutation. These 
sites are then grouped into maximally connected clusters using the algorithm introduced by Hoshen and Kopelman \cite{Hoshen1976a}.
 The identification of clusters depends on the choice of boundary conditions (periodic or non periodic). 
For the case of bulk germs, we use periodic boundary conditions. For surface germs, periodic boundary conditions are used only along the directions perpendicular to the incident wavevector (see section~\ref{sec:SurfaceGerms} for details).

\medskip
Once the clusters are identified, we compute  two quantities pertaining to the full distribution of clusters,
namely $\xi$, the correlation length,  and $P(s)$, the distribution of $s$, the cluster size. These quantities will be used in our analysis of the structure factor of both single clusters and clusters assemblies. $P(s)$ is determined by counting the clusters in a small interval around each $s$ value. $\xi$, which measures the average extent of non-percolating clusters, is defined as usual \cite{Stauffer2018}:

 \begin{equation}
\xi^2=\frac{\sum_{\rm{clusters}} 2 s^2 R_{g}^2(s)}{\sum_{\rm{clusters}}  s^2 }
\label{eq:xi}
\end{equation}

where the sum bears over all Bernoulli clusters, \textit{excepting the percolating one} (when it exists, \textit{i.e.} above $p_c$), and  $R_{g}$ is the cluster radius of gyration,
given by:
\begin{equation}
R_{g}^{2}=\frac{1}{s}\sum_{i}\left|\textbf{X}_{i}-\textbf{M}\right|^{2}
\end{equation}
where $\textbf{X}_i$ is the position of the $i^{th}$ site, $\textbf{M}$ the center of mass of 
the cluster and $s$ its number of sites. When $p$ increases from $p_c$ to 1, $\xi$ decreases from $\infty$ down to the mesh size .

\medskip
Values of $\xi (p)$ and plots of of $\xi (p)$ and $P(s,p)$ obtained from our simulations are given in Appendix A. Below and close to $p_c$, the behavior of $\xi (p)$ agrees with the known scaling law $\xi (p) \propto |p/p_c-1|^{-\nu}$, with $\nu$=0.879~\cite{Ballesteros1999}. Also in agreement with litterature, $P(s,p_c)$ at the percolation threshold decays exponentially with size $s$ as $P(s,p_c) \propto s^{-\tau}$, with $\tau=1+d/d_f=2.189$ \cite{Lorenz1998}.

\subsection{Selection of Bernoulli clusters}
\label{sec:Selection}
In a second step, we select some of these clusters, according to one of the following rules :

\begin{itemize}
\item  largest cluster only
\item  clusters of fixed size $s$ (within some interval)
\item  clusters containing bulk germs at a given fraction.\\
These bulk germs are chosen at random within the $pV$ picked sites. Their fraction is referred to the total number of sites $V$, hence is smaller than $p$. In practice, fractions of 10$^{-7}$ to   10$^{-1}$ were used. This case corresponds to invasion from bulk germs.
\item  clusters contacting one of the six surfaces of the cubic sample.
 This case corresponds to invasion from the surfaces (see \S\ref{sec:SurfaceGerms} for details).
\end{itemize}

 \medskip
 In the following, we will call \textit{potential} the picked sites of density $p$, and \textit{active} the sites belonging to the selected clusters, denoting by $p_{\rm{eff}}$ their density. The active sites are a subset of the potential sites. In the problem of evaporation, the active sites at a given pressure correspond to pores effectively emptied at that pressure. In our simulations, $p_{\rm{eff}}$ is directly determined by counting the number of active sites for given selection rules.

 \subsection{Structure factor of selected Bernoulli clusters}
\label{sec:StructureFactorSelected}
In a last step, we compute the total scattering signal for the selected clusters. For a given scattering angle, this signal is given by  the structure factor at the corresponding transfer wavevector $\textbf{q}$, \textit{i.e.} by the modulus squared of the discrete Fourier transform: 

\begin{equation}
  \label{Eq:un}
   I(\textbf{Q}) = \frac{1}{V}
 \left|\sum_{\textbf{R}}n\left(\textbf{R}\right)e^{\imath\pi\textbf{R}\textbf{Q}}\right|^{2}
  = \frac{1}{V}\sum_{\textbf{R},\textbf{S}} n(\textbf{R})n(\textbf{S}) e^{\imath\pi\textbf{Q}(\textbf{R}-\textbf{S})}
\end{equation}
where the sum runs on all lattice sites $\textbf{R}$, $\textbf{Q}=\textbf{q}/\pi$, and  the site occupancy $n\left(\textbf{R}\right)=1$ if the site is active, and 0 otherwise.

\medskip
For periodic boundary conditions, the structure factor $I(\textbf{Q})$ is given by the Fourier transform to the all-to-all  pair correlation function
$C_1(\textbf{R})$, defined as the probability that a site
separated by $\textbf{R}$ from a site in state 1 
is also in state 1, \textit{independently} of the cluster to which it belongs. It is important to realize that, for assemblies of clusters,  $C_1(\textbf{R})$ differs from the usual correlation function $C(\textbf{R})$, defined as the
probability that a site 
separated by $\textbf{R}$ from a site in state 1 
belongs to the \textit{same} cluster. Hence, while some properties of $C(\textbf{R})$ are known from the scaling theory of percolation \cite{Stauffer2018}, this is not the case for $C_1(\textbf{R})$.
The difference between these two functions is dramatically illustrated by the case where, at the percolation threshold, all Bernoulli clusters are considered. In this situation, $C(\textbf{R})$ decays with $R$ as $R^{2(d_{f}-d)}$\cite{Stauffer2018} with $d=3$ \footnote{Note that this decay is faster than the well known $R^{d_{f}-d}$ decay of the correlation function for the percolating cluster only. This difference is explained by the contributions of the smaller, non percolating, clusters.}.  In contrast, since the sites are chosen at random with probability $p$, $C_1(\textbf{R})$ is constant and equal to $p$. This difference between 
$C(\textbf{R})$ and $C_1(\textbf{R})$ explains why the structure factor cannot generally be derived from the scaling theory of percolation.

\medskip
While the structure factor is defined for 
any vector of the reciprocal space,
we restrict its calculation to the reciprocal lattice ($Q_x=2 i/L$, with $i$ an integer between 0 and $L$/2, and similarly with integers $j$ and $k$ for $Q_y$  and $Q_z$). This allows us to use the fast Fourier
transform algorithm FFTW \cite{fft} to evaluate $ I(\textbf{Q})$. 

\medskip
For a given realization, $ I(\textbf{Q})$ strongly fluctuates as a function of
      $\textbf{Q}$ due to the so-called speckle phenomenon. In an
      actual structure factor measurement, these fluctuations are washed out
      by using a detector covering many speckle spots around a given
      scattering direction. In our simulations, we obtain a similar averaging effect by averaging over different angular orientations of $\textbf{Q}$.
       In the case of bulk germs, the selection of
       clusters respects the rotational invariance, and we average over all angular orientations at constant modulus $Q$ to keep only the modulus dependence 
       (the case of surface germs is more subtle and will be discussed in section~\ref{sec:SurfaceGerms}).
In practice, all entries with $\sqrt{i^2+j^2+k^2}$ between $Q.L/2-1/2$ and $Q.L/2+1/2$
are averaged together to give the scattering signal at $I\left( Q \right)$. 
In a similar way,  an isotropized $C_1\left( R \right)$ is obtained by averaging over all $\textbf{R}$ with a modulus between $R-1/2$ and $R+1/2$. In the following, we will use the same
symbol for the scattering signal before or after averaging over angles.

\medskip
Due to statistical noise, $ I(Q)$ and $C_1\left( R \right)$ depend on the realization. However, because we consider self-averaging quantities, and the lattice size is large,  the fluctuations from sample to sample are small. For $L=1024$, averaging over 4 samples was found enough to obtain precise results.

\medskip
In the simple Bernoulli percolation problem, all clusters are selected, and the sites are
randomly in state 0 or 1. A simple analytical calculation then yields  
$I(Q\ne 0) = p(1-p)$ and $C_1(R \ne 0)= p$. 
In the case of clusters selection, the same expressions, with $p$ replaced by $p_{\rm{eff}}$, 
 would give the structure factor and the correlation function \textit{if} the active sites were non correlated. 
In order to measure the coherence induced by the selection process,\textit{ i.e.} the excess (or loss) of scattering signal due to the non independent positions 
of the active sites, we will use
the normalized quantities $I_N(Q)$ and  $C_{1N} (R)$, obtained by dividing the absolute structure factor $I(Q)$ and correlation function $C_1 (R)$ by $p_{\rm{eff}} (1-p_{\rm{eff}})$ and 
$p_{\rm{eff}}$.  

\medskip
When the density of selected clusters is small, we can expect the intercorrelation terms to be negligible, and $C_1(\textbf{R})$ to be simply related to $C(\textbf{R})$. 
In Appendix B, we confirm this expectation in the case of size selection by directly computing  $C(R)$ for clusters of  given size $s$, and comparing it to $C_1(\textbf{R})$, obtained by the inverse Fourier transform of $I(Q)$ for an assembly of clusters of size $s$.
We find that $C(R)$  and $C_1(R)$ only differ through an additive constant. This shows that  $C(\textbf{R})$ can be computed from $I(Q)$, at a much smaller numerical cost than in the direct calculation ($p_{\rm{eff}} V \, log(V)$ versus $(p_{\rm{eff}}V)^2$ steps).
In \S\ref{sec:SizeSelection}, this will allow us  to generalize to a wider range of $s$ and $p$ values an original result of our direct calculation of $C(R)$, which is that $C(\textbf{R})$ decays slower than exponential at large $R$. 

\section{Single clusters}
\label{sec:oneCluster}
In this section, we describe the $p$ dependence of the structure factor of, first, the largest cluster,  then of clusters of given size $s$. Beyond confirming the well-known fractal structure of percolation clusters \cite{Stauffer2018}, our results bring new informations concerning the quantitative $Q$ dependence of the structure factor, as well as the range of validity of the fractal structure away from the percolation threshold $p_c$.  

\subsection{Largest cluster}
\label{sec:largestCluster}

\begin{figure}[ht]
 \includegraphics[scale=0.3]{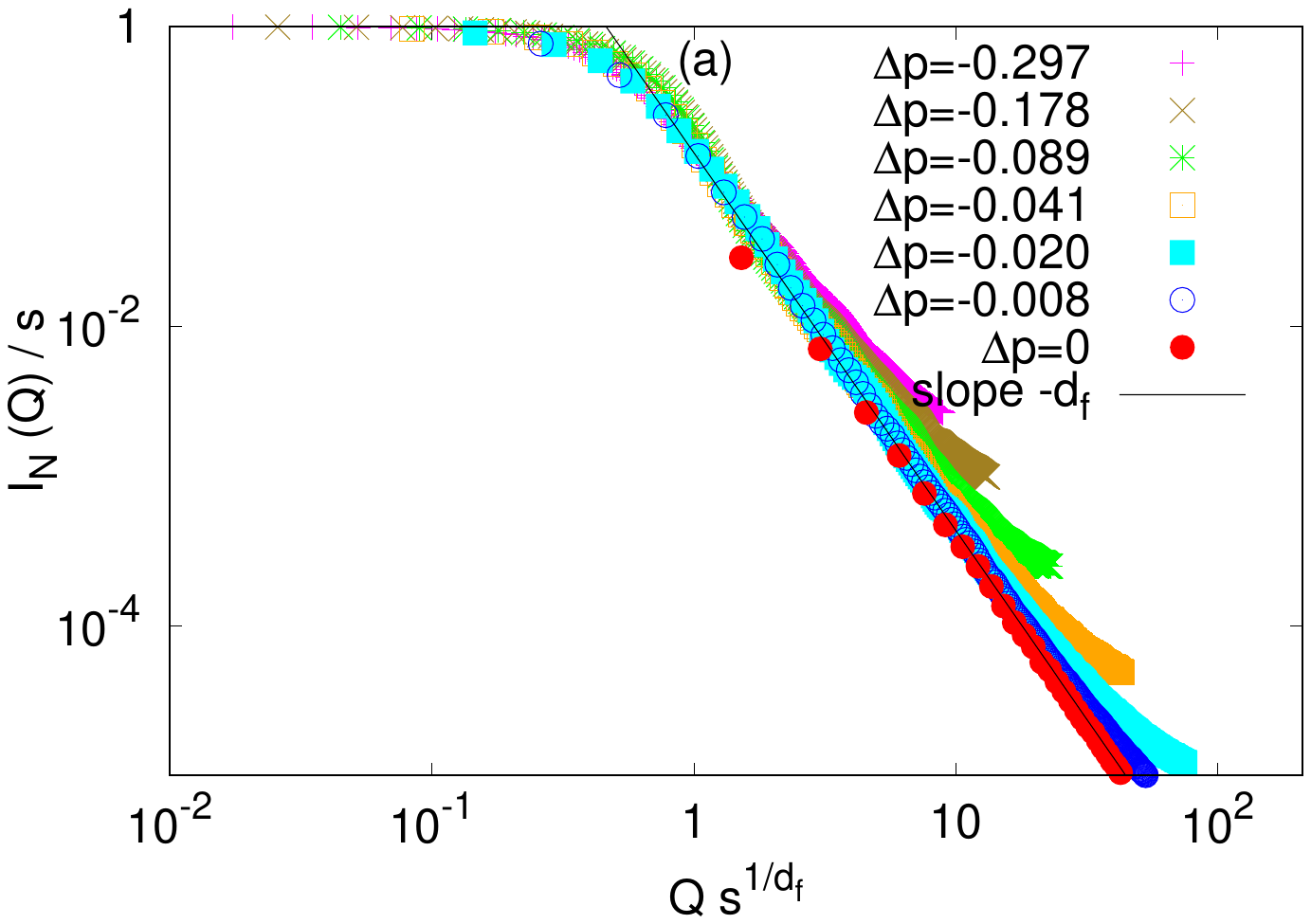}
 \includegraphics[scale=0.3]{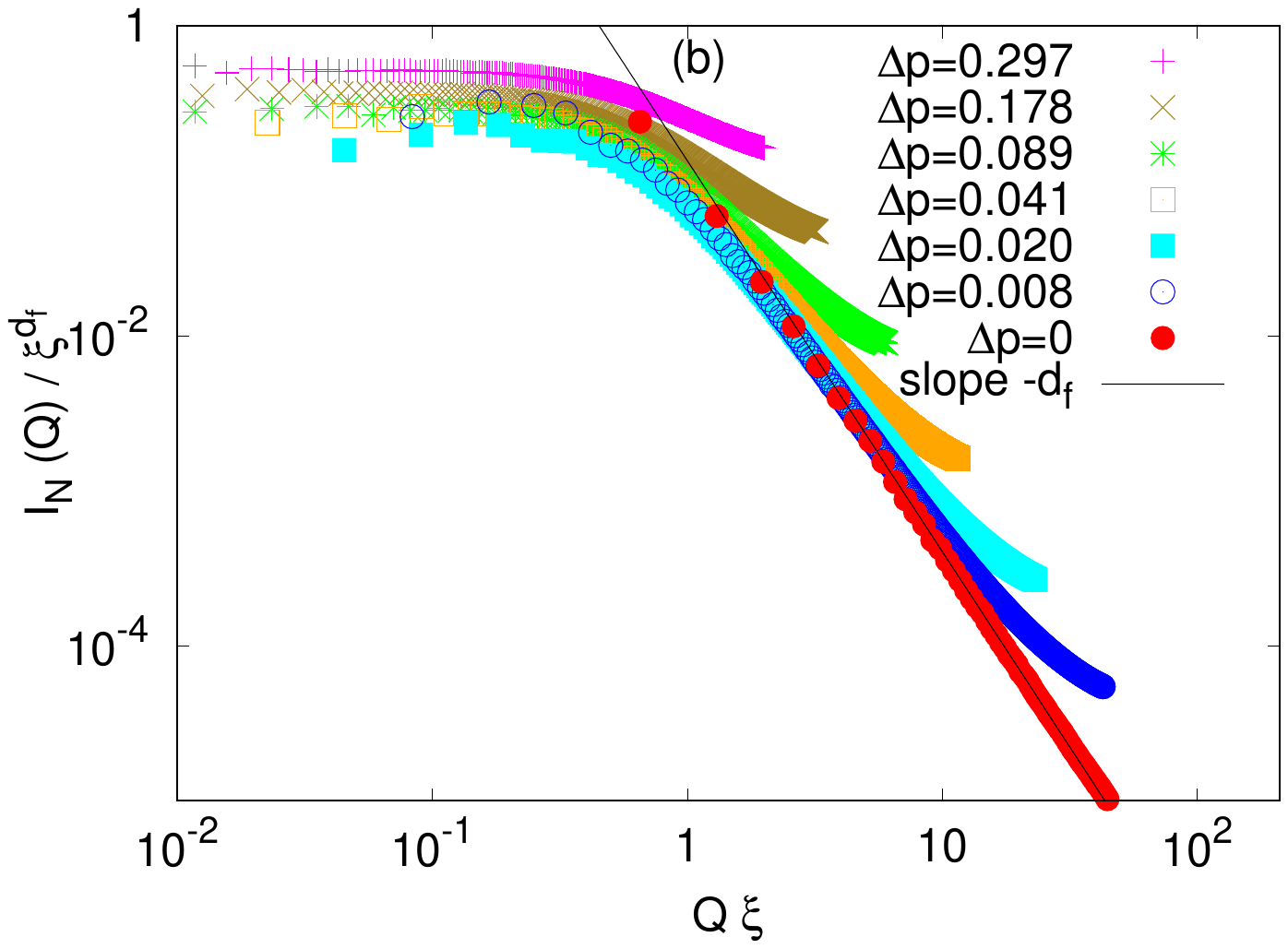}
  \caption{Normalized structure factor $I_N(Q)$ of the largest cluster for a 1024$^{3}$ lattice: (a) below $p_c$; (b) above $p_c$. Labels give $\Delta p = (p-p_c)/p_c$. Below $p_c$, the $Q$ scale is normalized by $s^{-1/d_f}$, where $s(p)$ is the size of the largest cluster. Above $p_c$, the $Q$ scale is normalized by $1/\xi$, where $\xi$ is the correlation length computed for the full cluster distribution, excluding the percolating cluster.
}
  \label{fig:SOLargest}
\end{figure}

 Figures~\ref{fig:SOLargest}a and b respectively show $I_N\left( Q \right)$ 
for the largest cluster,  for various $p$ below and above $p_c$. 

\medskip
Below $p_c$, the structure factor data collapse when the $Q$ scale is normalized by $s^{-1/d_f}$, where $s(p)$ is the cluster size, and $I_N\left( Q \right)$ is normalized by $s$. For large $Q s^{1/d_f}$ values, $I_N\left( Q \right)$ decreases as $Q^{-d_f}$, with $d_f \simeq 2.53$. This is in agreement with  the fractal structure of this cluster in a range of scales, with a fractal dimension characteristic of the 3D percolation problem. At small $Q$'s, the fractal range is limited by the cluster transverse extent $s^{1/d_f}$, so that  $I_N\left( Q \to 0 \right)/s $ saturates around 1 for $Q s^{1/d_f} \approx 1$. At larger $Q$'s, deviations from the $Q^{-d_f}$ behavior are observed for $Q\gtrsim 0.2$, due to the breakdown of fractality at the scale of the mesh size.
 
\medskip
Below $p_c$, $I_N\left( Q \to 0 \right)$ increases with $p$ because $s(p)$ increases. In contrast, above  $p_c$, while $s(p)$ keeps on increasing, $I_N\left( Q \to 0 \right)$ decreases,  This decrease follows from the progressive incorporation, above $p_c$,
 of clusters of intermediate size into the percolation cluster, which makes this cluster more homogeneous at smaller and smaller sizes. As noted by Stauffer and Aharony \cite{Stauffer2018}, one can expect the larger holes in the percolating cluster to be of order of the typical extent of the other clusters. This typical extent is expected to be of order the correlation length $\xi$.
Accordingly, we normalize in Fig.~\ref{fig:SOLargest}b the $Q$ scale  by $1/\xi$, computed from our simulations, and  $I_N\left( Q \right)$ by $\xi^{d_f}$. The approximate collapse of data 
shows that, above $p_c$, the percolation cluster is homogeneous at scales larger than $\xi$, and close to a fractal of dimension $d_f$ at smaller scales, in agreement with Stauffer and Aharony's expectation.

\subsection{Clusters of fixed size $s$}
\label{sec:SizeSelection}

\begin{figure}[ht]
\includegraphics[scale=0.3]{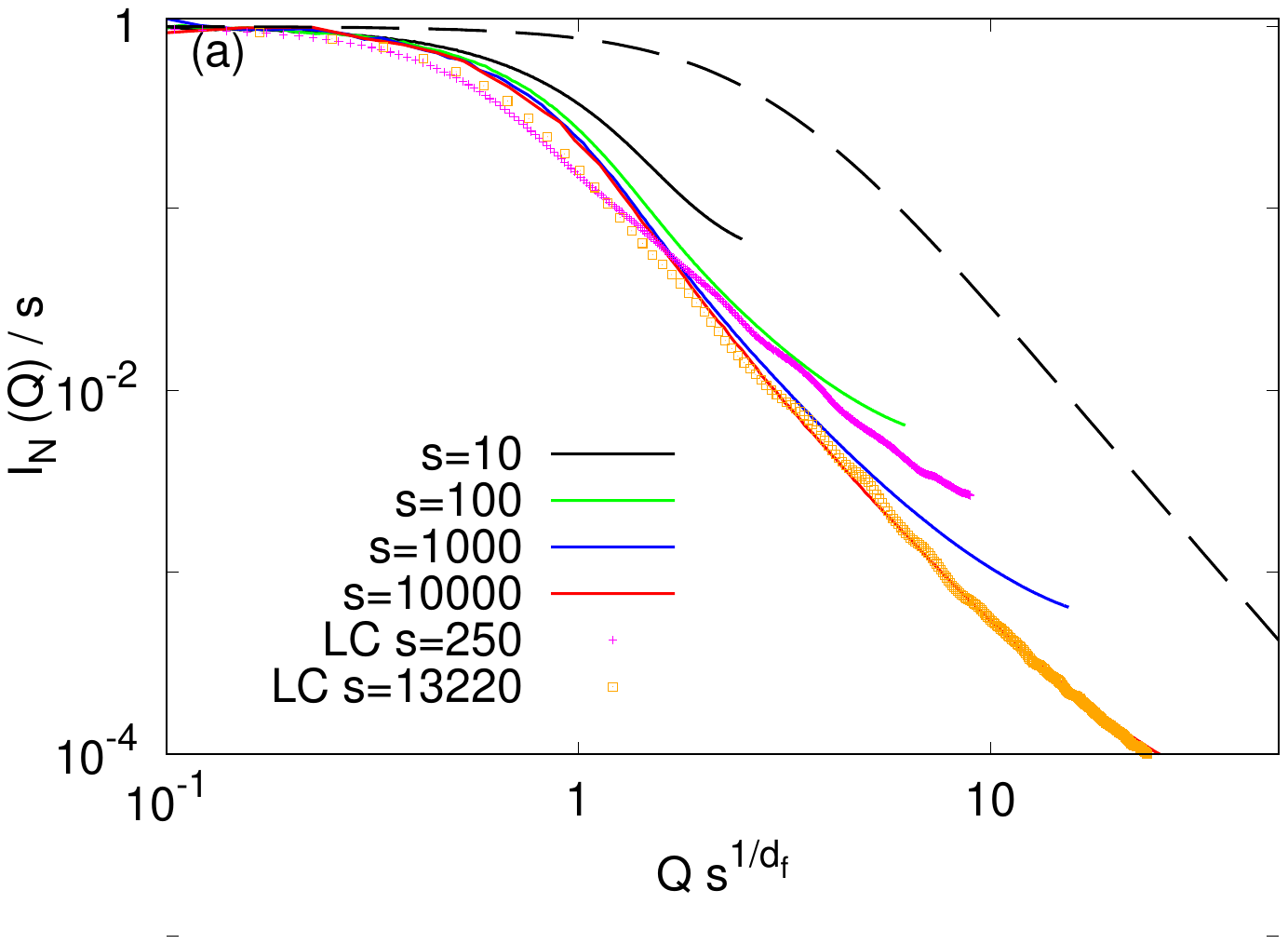}
\includegraphics[scale=0.3]{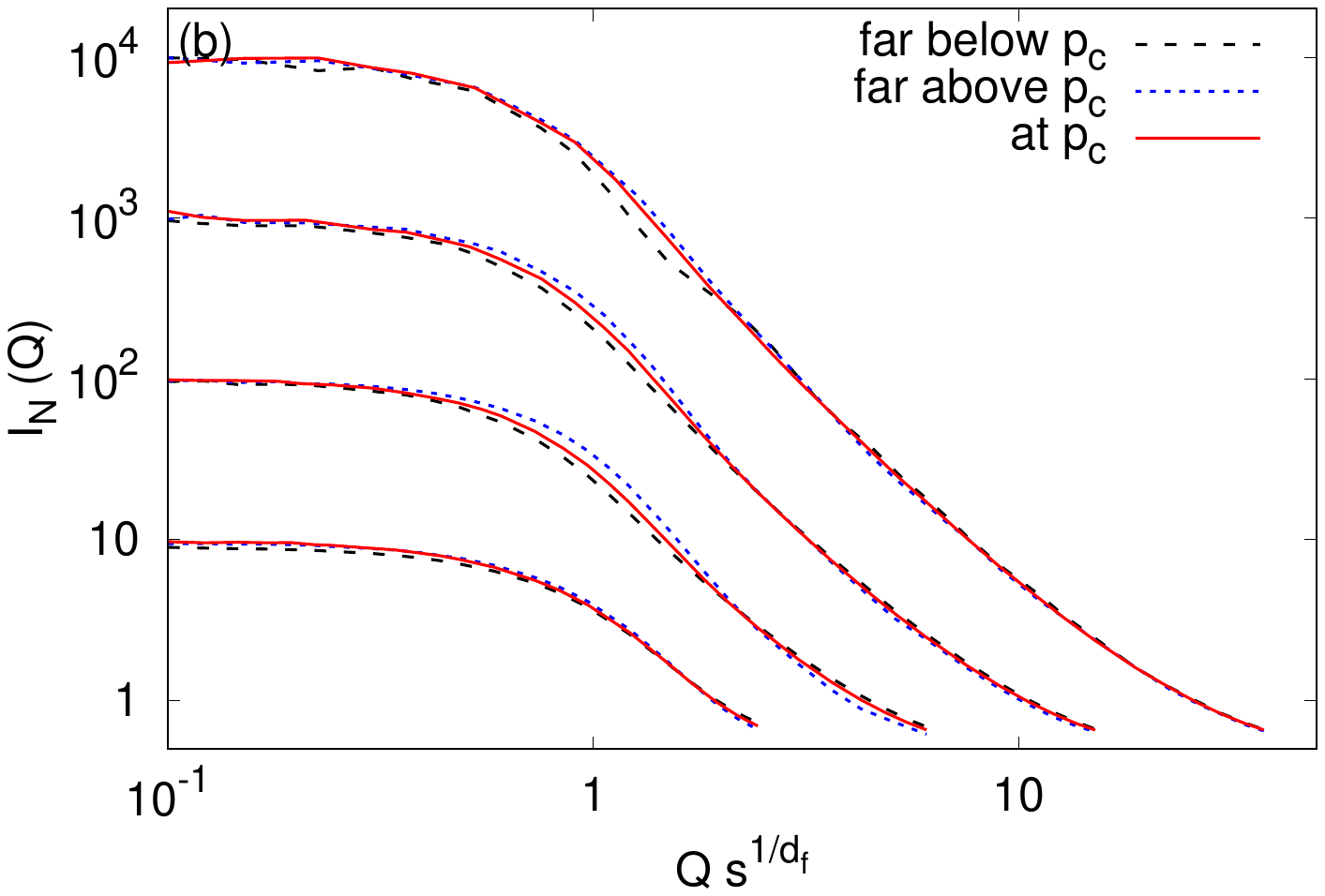}
 \caption{Normalized structure factor $I_N(Q)$ of individual clusters of size $s$. $I_N(Q)$ is computed by selecting all clusters of size of size $s$ within an interval $\pm 5\%$. The axes are normalized  as in Fig.~\ref{fig:SOLargest}a;  (a) Scaling behavior at percolation for $s$=10, 100, 1000 and 10000. The structure factor of the largest cluster is also shown for similar sizes (250 corresponding to -30\% below $p_c$ and 13000 corresponding to -4\% below $p_c$). The dashed curve is the theoretical structure factor if  the correlation function decays exponentially with $R$ (Eq.~\ref{eq:Sinha}); (b) Comparison of the structure factor for the same sizes $s$ ($s \approx I_N(Q \to 0$)) away from percolation .
 For each size, the structure factor is shown at $p_c$ and for the smallest and largest values of $p$ giving clusters of that size ($\Delta p= \pm$ 30\% for sizes 10 and 100, $\pm$8\% for size 1000, and (-4\%, 2\%) for size 10000). 
 In this range of $p$, the structure of clusters appears essentially controlled by their size, independently of the distance to the percolation threshold.}
  \label{fig:SO_scaled_1024_3D}
\end{figure}

We now consider the case where only clusters of given size $s$ within some interval $\Delta s$ are selected. The resulting structure factor at $p_c$ is represented in Fig.~\ref{fig:SO_scaled_1024_3D}a, normalizing the 
axes as in Fig.~\ref{fig:SOLargest}a.
 The different curves correspond to  $s$ increasing from 10 to 10000 by steps of 10, and a constant relative interval $\frac{\Delta s}{s}=\pm 5\%$. 
  In all cases, the average distance between clusters is large compared to their spatial extent.
    We  then expect the structure factors of the different clusters to add incoherently,
    and the average structure factor \textit{per} cluster to give the average structure factor for a cluster of given size $s$.
    In Appendix B, we check this expectation by showing that, within a constant, the all-to-all correlation function $C_1$
      for the above selection coincides with $g_s(R)$, the average of the correlation function
       $C(R)$ over all clusters of size $s$ within $\Delta s$. 

\medskip
The good collapse of data observed in Fig.~\ref{fig:SO_scaled_1024_3D}a is consistent with the known property that,
at percolation, all clusters have a fractal structure with exponent $d_f$, extending from the mesh size up to their extent.
Away from the percolation point, scale invariance implies that clusters keep a fractal structure,
but with a fractal exponent dependent on $p$ \cite{Stauffer2018}.
 To the best of our knowledge, this dependence has not been yet quantified. Our simulations of the structure factor of the largest cluster  (Fig.~\ref{fig:SOLargest}) suggest  that it is weak in a range $\pm$ 30\% around $p_c$.
  This conclusion is consistent with the comparison, shown in Fig.~\ref{fig:SO_scaled_1024_3D}a, of the structure factor of clusters of given size at $p_c$ to that of the largest cluster at $p$ values chosen to give similar sizes. 
The same comparison could suggest a possible dependence with $p$ of the  $Q$ dependence around $(s)^{-1/d_f}$,
 but this dependence may also be an effect of the small statistics ($I_N(Q)$ for the largest cluster is averaged over four realizations only).
 In order to clarify this point, we compare in Fig.~\ref{fig:SO_scaled_1024_3D}b, the average structure factors for the same cluster sizes, below, at, and above $p_c$.
  For each size, the structure factor is shown at $p_c$ and for the smallest and largest values of $p$ giving clusters of that size.
  The near collapse of data shows that the fractal structure of clusters barely depends to the distance to the percolation threshold in the probed range of $p$ values. This range is however limited by the finite size of our sample, as large size clusters are very unlikely too far below $p_c$. 
In order to extend the probed range of ($p$, $s$) couples of values, we have used an invasion algorithm to generate large clusters away from $p_c$. The results, described in Appendix B, confirm that the structure factor of individual clusters of a given size is only weakly sensitive to $p$ in a wide range below $p_c$. 

\medskip
 To our knowledge, this remarkable feature has not yet been previously reported in the litterature. 
In practice, we will use this property in Appendix C to explain the $Q$ dependence of the scattering signal for invasion from bulk germs, using a simple parametrization of the simulated $Q$ dependence of $I_N(Q)$.

\medskip
Figure~\ref{fig:SO_scaled_1024_3D}a also provides a quantitative description of the $Q$ dependence of the structure factor $I_N(Q,s)$. Noteworthingly, $I_N(Q,s)$ for percolation clusters differs from the expression widely used for fractal colloids \cite{Freltoft1986a,Ferri1991a}. 

\begin{equation}
 I_N(q,s)/s =\frac{\sin[(d_f-1) \arctan(q\lambda)] }{(d_f-1)q\lambda} \frac{1 }{(1+(q\lambda)^2)^{(d_f-1)/2}}
\label{eq:Sinha}
 \end{equation} 
with $\lambda = (\frac{s}{4\pi \Gamma (d_f)})^{1/d_f} \approx 0.327 s^{1/d_f} $  and $\Gamma(x)$ the gamma function.

\medskip
As shown by the dashed line in Fig.~\ref{fig:SO_scaled_1024_3D}a, the crossover from the low $Q$ to the large $Q$ behavior occurs in the latter case at significantly larger wavevectors.
This difference follows from the fact that Eq.~\ref{eq:Sinha} assumes
the two-points correlation function $g_s(r)$ to behave as $g_s(r) \sim r ^{d_f-3} exp (-r/\lambda)$ \cite{Sinha1989a}.
In contrast, our direct calculation in specific cases of the two points correlation function of Bernoulli clusters (see Appendix~B) shows that the initial decay at small $r$'s of the scaling function $g_s(r) r^{3-d_f
}$ is slower than the above exponential. The near collapse of the different $I_N(Q,s)$ in Fig.~\ref{fig:SO_scaled_1024_3D}a allows to generalize this conclusion to a wider range of $s$ and $p$ values than used in Appendix~B.

\section{Bulk germs}
\label{sec:BulkGerms}

\begin{figure}[ht]
  \includegraphics[scale=0.55]{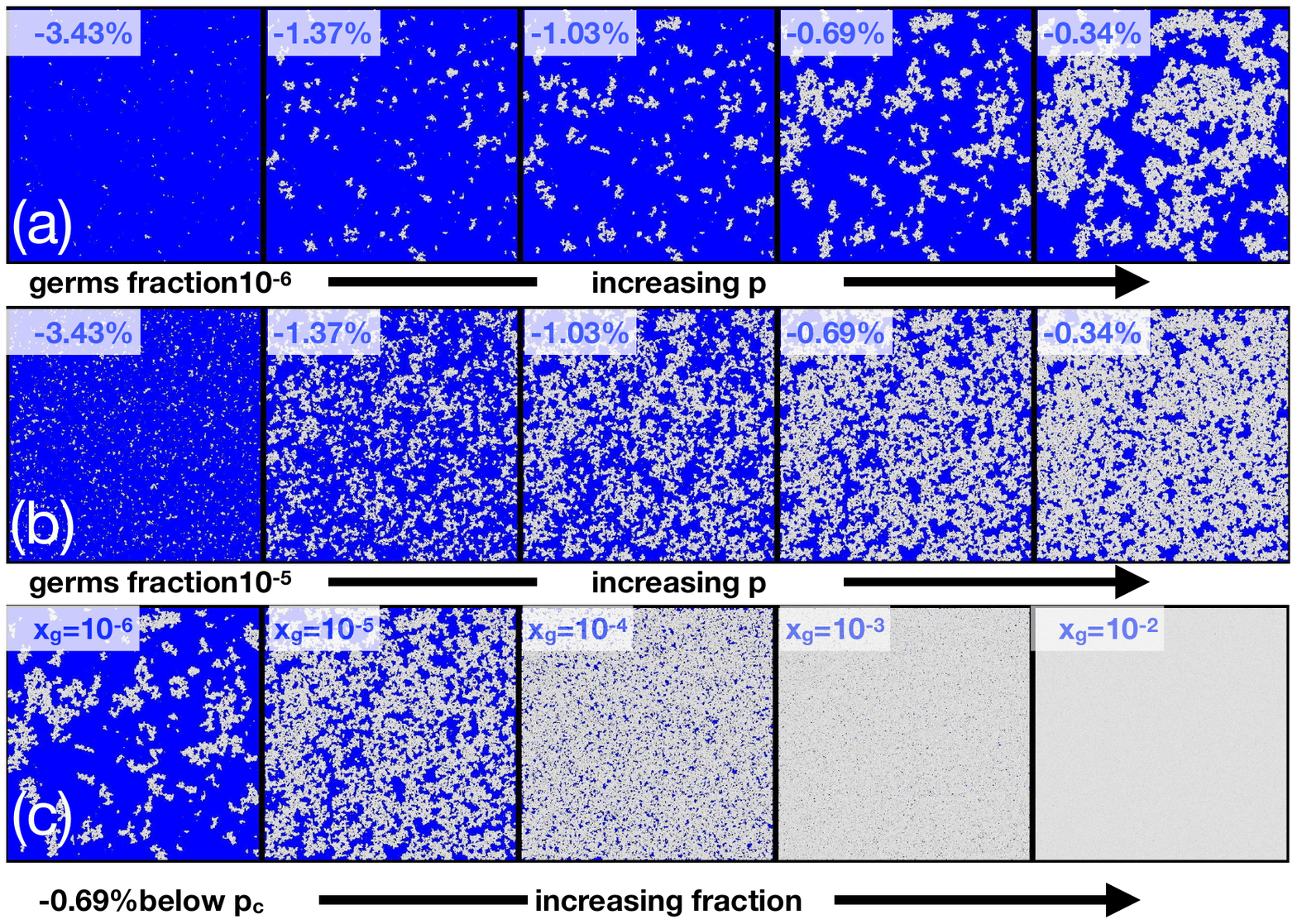}
  \caption{Evolution of the selected clusters when the density of germs, 
or the distance to the percolation threshold, varies. In these 2D simulations 
for a 16384$^2$ lattice, regions of 32x32 pixels were averaged to yield  
512x512 images. The grey level measures the number of selected sites in 
each metapixel. Note that the smallest clusters cannot be resolved. 
Row (a): fixed fraction of germs $10^{-6}$ (corresponding to about 
250 germs), relative distance to $p_c$ increasing from -3.4\% to -0.34\%.  
Raw (b): fixed fraction of germs $10^{-5}$ including germs of raw (a), same values of $p$. 
Raw (c): fixed distance to the percolation threshold (-0.69\%), 
fraction of germs increasing from $10^{-6}$ to $10^{-2}$.}
  \label{fig:pictures_clusters}
\end{figure}

\medskip
We now turn to the case where selected clusters are those containing germs randomly chosen among the potential sites. We will denote by $x_g$ the density 
of such germs, \textit{i.e. }their number divided by $V$, the total number  
of sites in the simulation volume. The scattering problem then involves three length scales: $\xi$, the correlation length for all non percolating clusters, which, below $p_c$, measures the typical extent of clusters,  $d_g=(x_g)^{-1/d}$, the characteristic distance between germs  at $d$ dimensions, and $1/Q$ which measures the scale probed by the radiation. The values of $\xi$ and $d_g$ for our 3D simulations are given in Appendix~A (Fig.~\ref{fig:xidg}).

\medskip
Figure~\ref{fig:pictures_clusters} illustrates the influence of the germ fraction $x_g$ for a 2D lattice, when the distance to
 percolation and the germ fraction are varied. In Figs.\ref{fig:pictures_clusters}a and b, $x_g$ is
  a constant, respectively equal to $10^{-6}$ and $10^{-5}$.
   As the percolation threshold is gradually approached,  the correlation length $\xi$
   increases, and the selected clusters grow in size. As long as $\xi$ remains much smaller than $d_g$, the clusters
    are mostly well separated. In this \textit{dilute} regime, we expect $I(Q)$  to be given by the addition of the structure factors of the individual clusters, and to increase with the fraction of germs. In contrast, when $\xi$ becomes comparable to or larger than $d_g$, we enter a \textit{concentrated} regime: 
    the clusters are closer and some of them merge, making their number smaller than the number of germs. Simultaneously, the sample becomes more homogeneous at large scales. Figure~\ref{fig:pictures_clusters}c shows a similar evolution when the germ fraction is increased at a constant distance to percolation. Hence, in the concentrated regime, we expect $I_N(Q)$  at small $Q$ to decrease when increasing the fraction of germs. 
In the following, we discuss these two regimes in the 3D case.

\subsection{Dilute regime below $p_c$}
\label{sec:Dilute}

\begin{figure}[ht]
\includegraphics[scale=0.3]{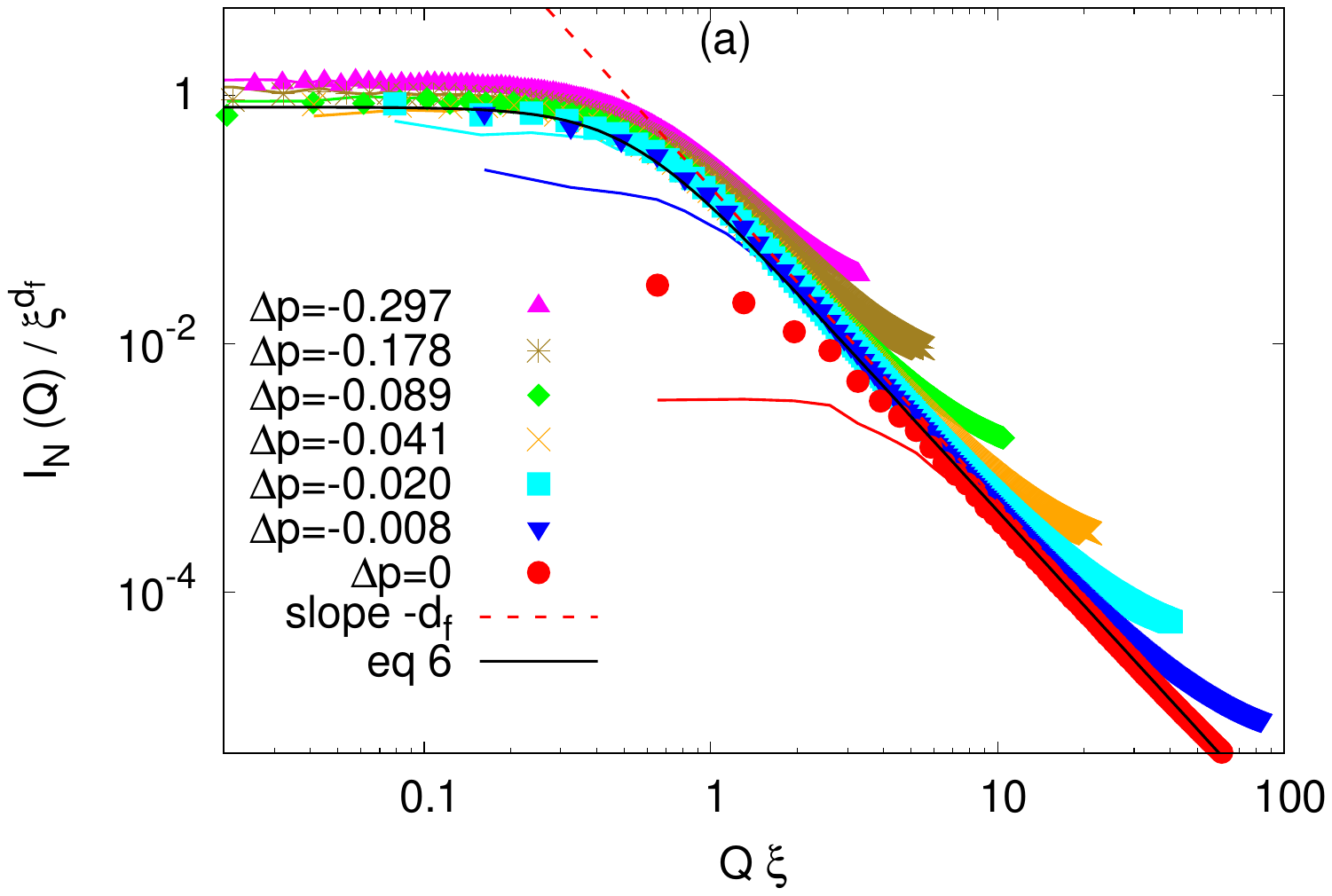}
\includegraphics[scale=0.3]{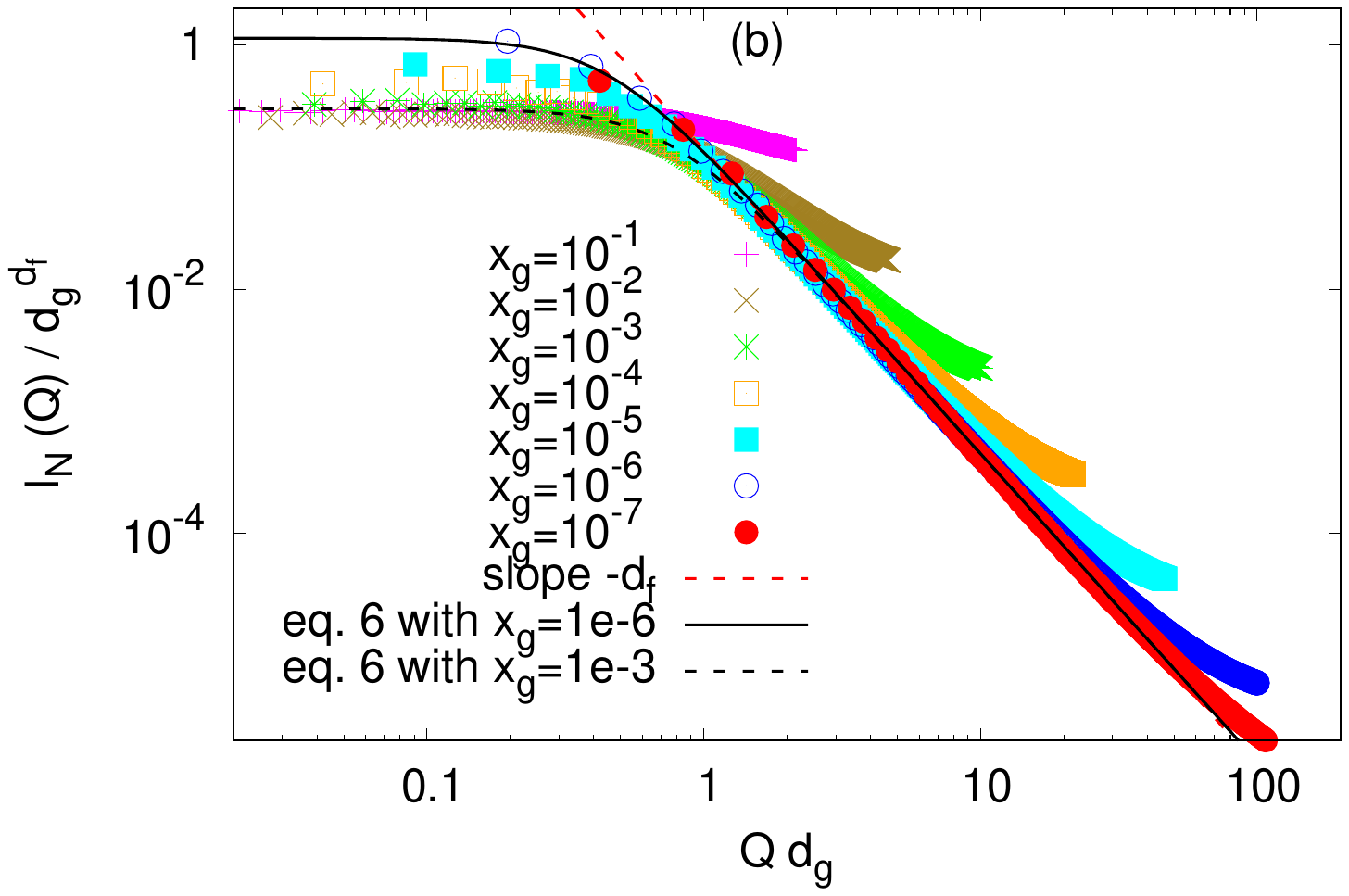}
\caption{Normalized structure factor $I_N\left(Q, \, p, x_g\right)$ below the percolation threshold for invasion from bulk germs;  (a):  Demonstration of  the dilute regime for two germ fractions, $x_g$= $10^{-6}$ (symbols) and $10^{-5}$ (lines),
and relative distances to $p_c$,  $\Delta p= (p-p_c)/p_c$,from -30\% to 0. 
The $Q$ scale is normalized by $1/\xi$, where $\xi$ is the correlation length computed for the full cluster distribution, and  $I_N\left(Q\right)$ by $\xi^{d_f}$. The continuous black line is the prediction of Eq.~\ref{INtheoQ}; (b): Demonstration of  the concentrated regime close to the percolation threshold($\Delta p$= -0.2 \%, $\xi\approx 300$) for $x_g$ increasing 
from $10^{-7}$ to $10^{-1}$. The $Q$ scale is normalized by $1/d_g$, where  $d_g$ is the characteristic distance between germs, and  $I_N\left(Q\right)$ by $d_g^{d_f}$. The continuous and dotted black lines are the predictions of Eq.~\ref{INtheoQ} for respectively $x_g$=10$^{-6}$ and 10$^{-3}$.
}
\label{fig:SO-Q-germs-belowpc}
\end{figure}

 \medskip
 Figure~\ref{fig:SO-Q-germs-belowpc}a shows the computed structure factor for two germs densities, $10^{-6}$ and $10^{-5}$, and different $p$ values below $p_c$.
  For such densities, the average distance between germs is larger than $\xi$, except very close to $p_c$, and we fall in the dilute regime.
  In this regime, we expect the normalized structure factor to only depend on $Q$ and $\xi$ and the fractal structure of individual clusters to be preserved.
   Accordingly, we normalize the axes of Fig.~\ref{fig:SO-Q-germs-belowpc}a by $1/\xi$ and $\xi^{d_f}$, respectively. The good collapse of data confirms our expectations. 
   As in the case of the largest cluster, the deviations at large values of $Q.\xi$ reflect the effect of the mesh size. 
   Deviations at small angles observed at $p_c$  for the two fractions, and slightly below $p_c$ for the largest one reflect the breakdown of the dilute regime, when $d_g$ becomes comparable to $\xi$.
 
 \medskip
Two important conclusions can be drawn from Fig.~\ref{fig:SO-Q-germs-belowpc}a. First, the measurements of the normalized structure factor at small $Q$  give access to $\xi$, once $d_f$ is determined. Second, $d_f$ can be measured from the decay of the structure factor at large $Q$ values, at or slightly close to $p_c$. This conclusion seems at variance with a previous study of  Martin and Ackerson \cite{Martin_PRA1985}, who have shown that, at the percolation point, the structure factor
    of a dilute assembly of clusters should decay with $Q$ with an exponent  $(\tau-3) d_f$=$3-2d_f$, different from $-d_f$.  In fact, this discrepancy comes from the difference between the cluster size distribution $P(s)$  in the context of gelation studied by Martin and Ackerson, and the distribution $N(s)$ relevant in the context of invasion from bulk germs. In the former case, where gelation clusters are separated apart by dilution, $P(s)$ is the size distribution function for Bernoulli clusters at percolation, which decays as $P(s) \propto s^{-\tau}$, with $\tau$ =2.189 at $d$=3. In the latter case,  $N(s)=s.P(s)$, as, when picking a germ at random among all potential sites,  the probability of selecting a cluster of size $s$ is proportional to $s$. As detailed in appendix C, the slower decay of $P(s)$ compared to $N(s)$ implies that the structure factor is controlled, in a similar way to $\xi$, by the largest clusters. This explains why the $Q^{-d_f}$ behavior is preserved by the average over clusters.

\begin{figure}[ht]
  \includegraphics[scale=0.5]{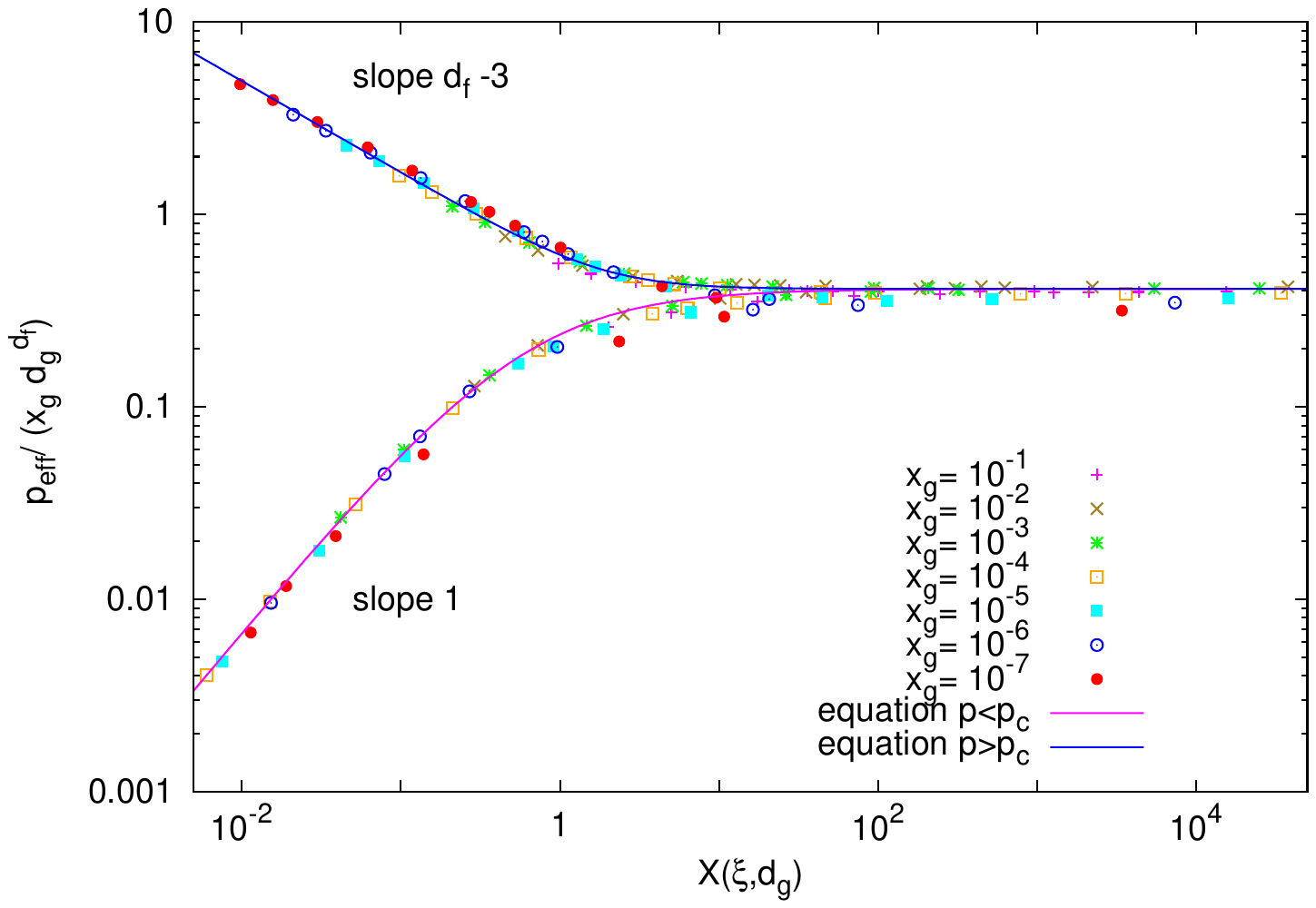}
 \caption{Scaling behavior of $p_{\rm{eff}}$, the density of active sites, for invasion from germs;
   a plot of $p_{\rm{eff}}/ (x_g d_{g}^{d_{f}})$ against $X$=$
   \xi^{(3-\tau) d_f}/d_{g}^{d_{f}}$ (below $p_c$)  and  $X$=$\xi/d_{g}$ (above
   $p_c$) gives a good collapse of all data. Close to $p_c$ (large $X$), the constant asymptotic behavior corresponds  to a dense assembly of fractals blobs of density $x_g$ and average extent $d_g$,
    $p_{\rm{eff}} \approx x_{g} d_{g}^{d_f}$. Away from $p_c$ (small $X$), the asymptotic power laws correspond to 
     $p_{\rm{eff}} \approx x_g \xi^{(3-\tau) d_f}$ below $p_c$ (dilute clusters) and
      $\xi^{(d_f-3)}$ above $p_c$ (concentrated regime dominated by the percolating cluster for $\xi \ll d_{g}$).
   The lines correspond to Eqs.~\ref{pefftheoavantaprespc}.}
\label{fig:scalingpeff}
\end{figure}

 \medskip
Notwithstanding, the cluster size distribution comes into play  when considering, rather than the normalized structure factor, the physically measured quantity, the \textit{absolute} structure factor. Indeed, $I(Q)$= $p_{\rm eff} (1-p_{\rm eff}) I_N(Q)$, involves  $P(s)$  through $p_{\rm eff}$. As shown by our simulations (Fig.~\ref{fig:scalingpeff}) and explained in appendix C,  $p_{\rm eff} $ scales in the dilute regime as $\xi^{(3-\tau) d_f}$.
This implies that the low $Q$ absolute structure factor  diverges as  $\xi^{(4-\tau) d_f}$, hence depends on $\tau$.

\subsection{Concentrated regime below $p_c$ }
\label{sec:Concentrated}
We now discuss the behavior below $p_c$  for $d_g <\xi$, i.e.  in the concentrated limit. In this regime, we might expect the fractal structure of clusters to be limited by $d_g$ rather than $\xi$, hence the distribution of selected sites to consist in a dense assembly of fractals blobs of density $x_g$, average extent $d_g$, 
and fractal dimension $d_f$. As shown by Fig.~\ref{fig:scalingpeff}, the density $p_{\rm eff}$ of selected sites agrees with this expectation, varying as  $p_{\rm eff} \propto x_g \,d_g^{d_f}$. Accordingly, we plot in Fig.~\ref{fig:SO-Q-germs-belowpc}b the normalized structure factor, divided by $d_g^{d_f}$,  as a function of $Q d_g$. This scaling approximately collapses the curves obtained 
for increasing germ fractions at a  fixed distance to $p_c$, confirming that, in the concentrated regime, $d_g$ is the characteristic length of the scattering problem. At scales larger than $d_g$, the structure factor is constant, and the sample homogeneous. At smaller scales, the fractal structure is retained and the structure factor  varies as $Q^{-d_f}$. Overall, our simulations thus confirm that the spatial distribution of selected sites follows the blob picture.  In this regime, inter-cluster correlations are found to slightly modify the structure factor at small $Q$: Instead of $I_N(Q \to 0) \propto  \, d_g^{d_f}$, we indeed find  $I_N(Q \to 0) \approx  0.16 \, d_g^{d_1}$ with $d_1$=2.8, approximately 10\% larger than $d_f$. 

 \subsection{Behavior above $p_c$ }
\label{sec:Abovepc}

\begin{figure}[ht]
\includegraphics[scale=0.3]{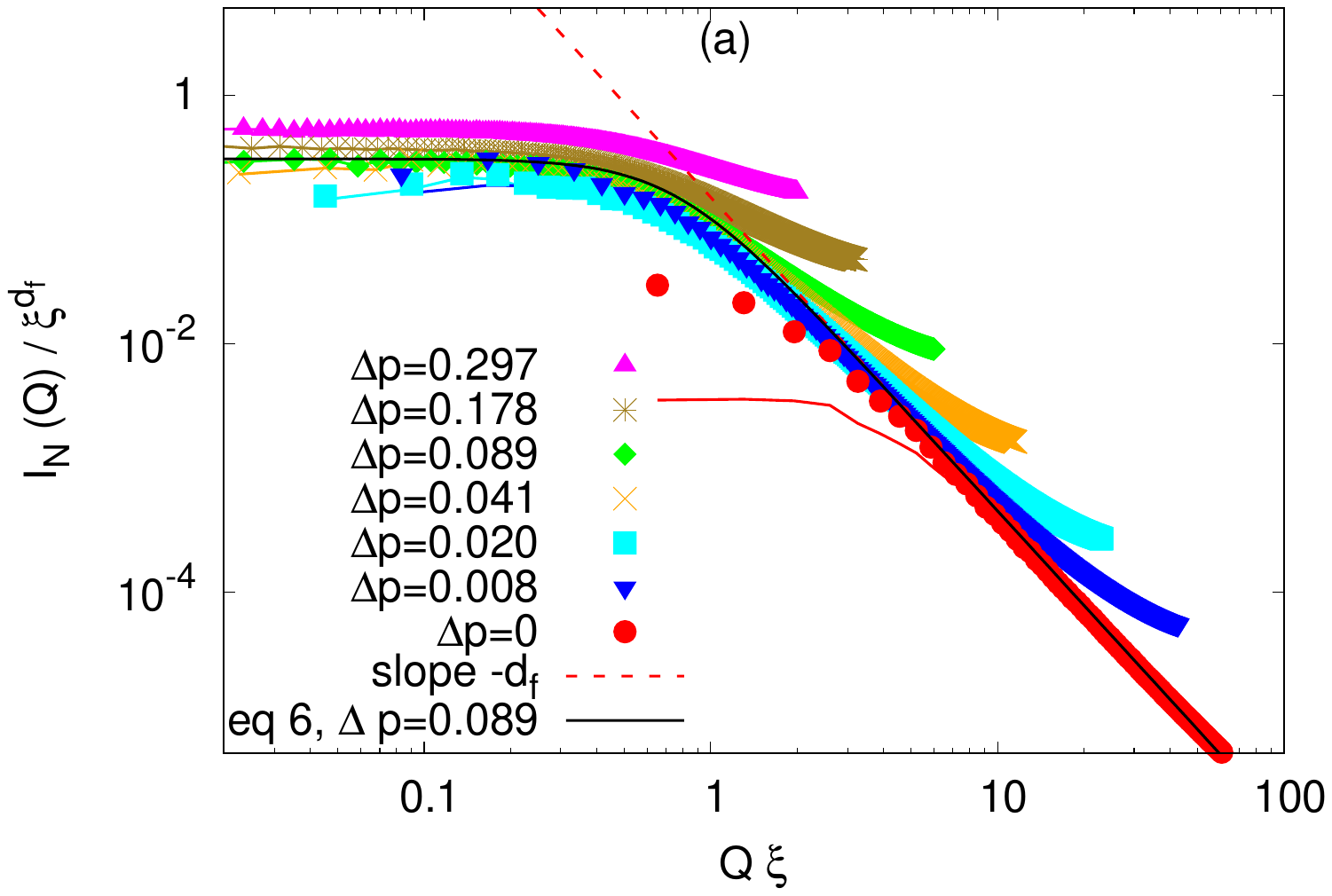}
\includegraphics[scale=0.3]{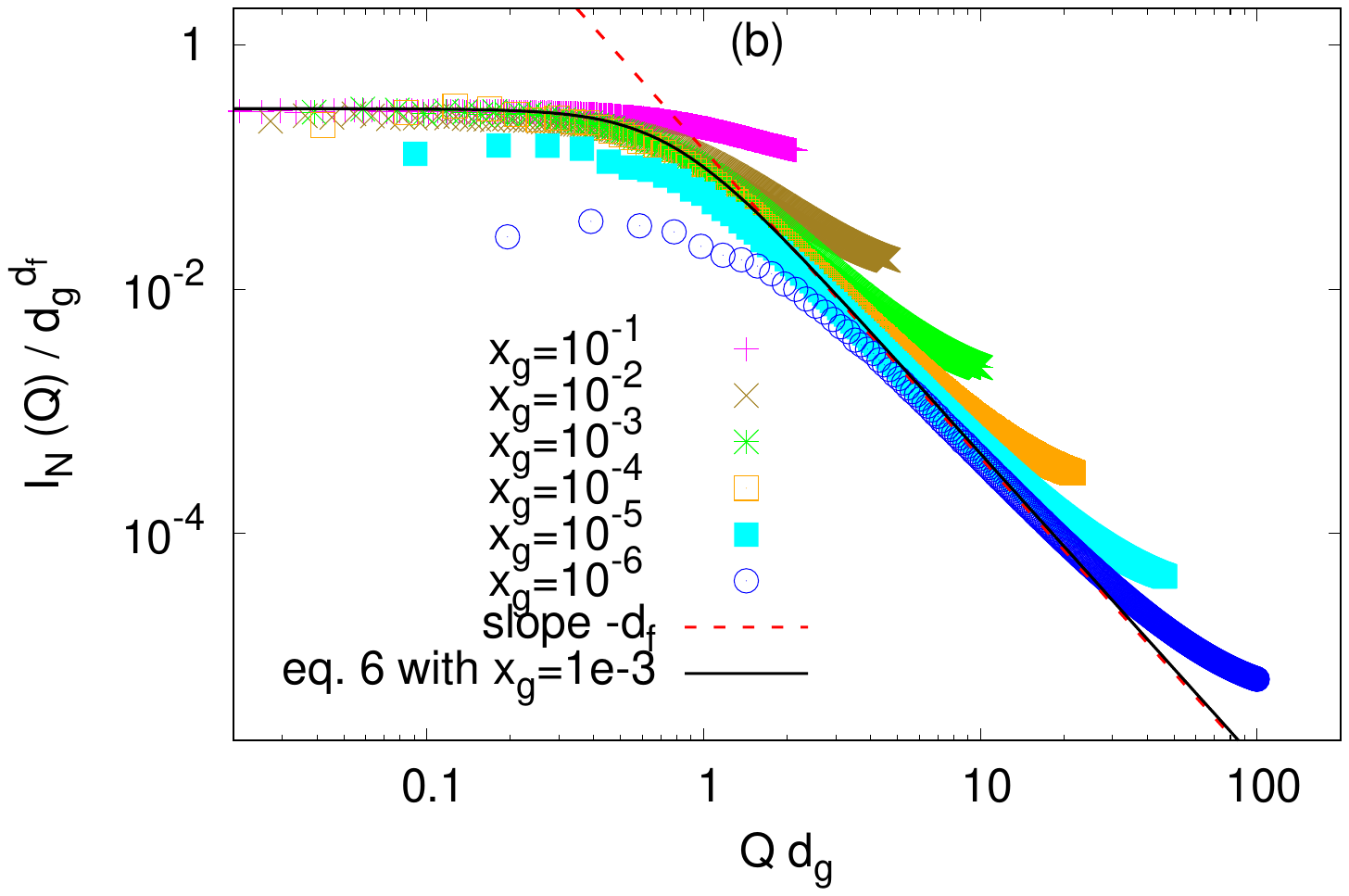}
\caption{Normalized structure factor $I_N\left(Q, \, p, x_g\right)$ of
clusters above the percolation threshold, for different distances to $p_c$ and germ fractions. 
Same axes as in Fig.~\ref{fig:SO-Q-germs-belowpc};
(a): for germ fractions $x_g$= $10^{-6}$ (symbols) and $10^{-5}$ (lines), and  $\Delta p$ ranging from 0 to 30\%. The continuous black line is the prediction of Eq.~\ref{INtheoQ} for  $\Delta p$=0.089 ($\xi \approx$ 6).
(b): for germ fractions increasing from $10^{-6}$ to
$10^{-1}$ and $\Delta p$ =0.8 \% ($\xi\approx 40$).
The continuous black line is the prediction of Eq.~\ref{INtheoQ} for  $x_g$=10$^{-3}$.
}
\label{fig:SdeQaprespc}
\end{figure}

Because the percolation transition is continuous, we expect the above results to also apply right above $p_c$. 
For a fixed germ fraction, two changes occur as $p$ further increases. First, preexisting disconnected clusters can connect
 into larger clusters, and, second, new clusters can be selected or preexisting clusters can abruptly grow in size by connection to another cluster previouly unselected. 
  Since the structure factor is
   unsensitive to the fact that selected sites belong or not to the same cluster,
   the first process  barely affects $p_{\rm eff}$ and  $I_N(Q \to 0)$.
A theoretical evaluation of the second process is not straightforward.  
   Fig.~\ref{fig:SdeQaprespc}b and Fig.~\ref{fig:scalingpeff} respectively show that, above  $p_c$, $I_N(Q \to 0)$ and  $p_{\rm eff}$ behave similarly to right below $p_c$ as long as $ \xi >d_g $. Namely, $I_N(Q \to 0) \propto d_g^{d_1}$, and $p_{\rm eff}\propto x_g.d_g^{d_f} \sim d_g^{d_f-3}$.

\medskip
In the opposite situation $\xi\ll d_g$ (i.e.  far above $p_c$), Fig.~\ref{fig:SdeQaprespc}a shows a behavior similar to that observed in the dilute regime in Fig.~\ref{fig:SO-Q-germs-belowpc}a,  $I_N ( Q)$ varying approximately as $\xi^{d_f}$ at low $Q \xi$, and decreasing as $Q^{-d_f}$ at large $Q $ ($<0.2$). This reflects the fact that, for small $\xi$, most selected sites belong to the percolating cluster, for which such a dependence was observed in Fig.~\ref{fig:SOLargest}b. Consistently, Fig.~\ref{fig:scalingpeff} shows that $p_{\rm eff}$ varies as $ \xi^{d_f-3}$, the density of the  percolating cluster. intercorrelation of the percolation cluster with the other clusters only slightly modify the total structure factor: $I_N(Q \to 0) $ is found to vary as $\xi ^{d_2}$, with $d_2=2.25 < d_f$.

\subsection{An unified description of scattering for selection by bulk germs}
\label{sec:Unified}
\begin{figure}[ht]
 \includegraphics[scale=0.3]{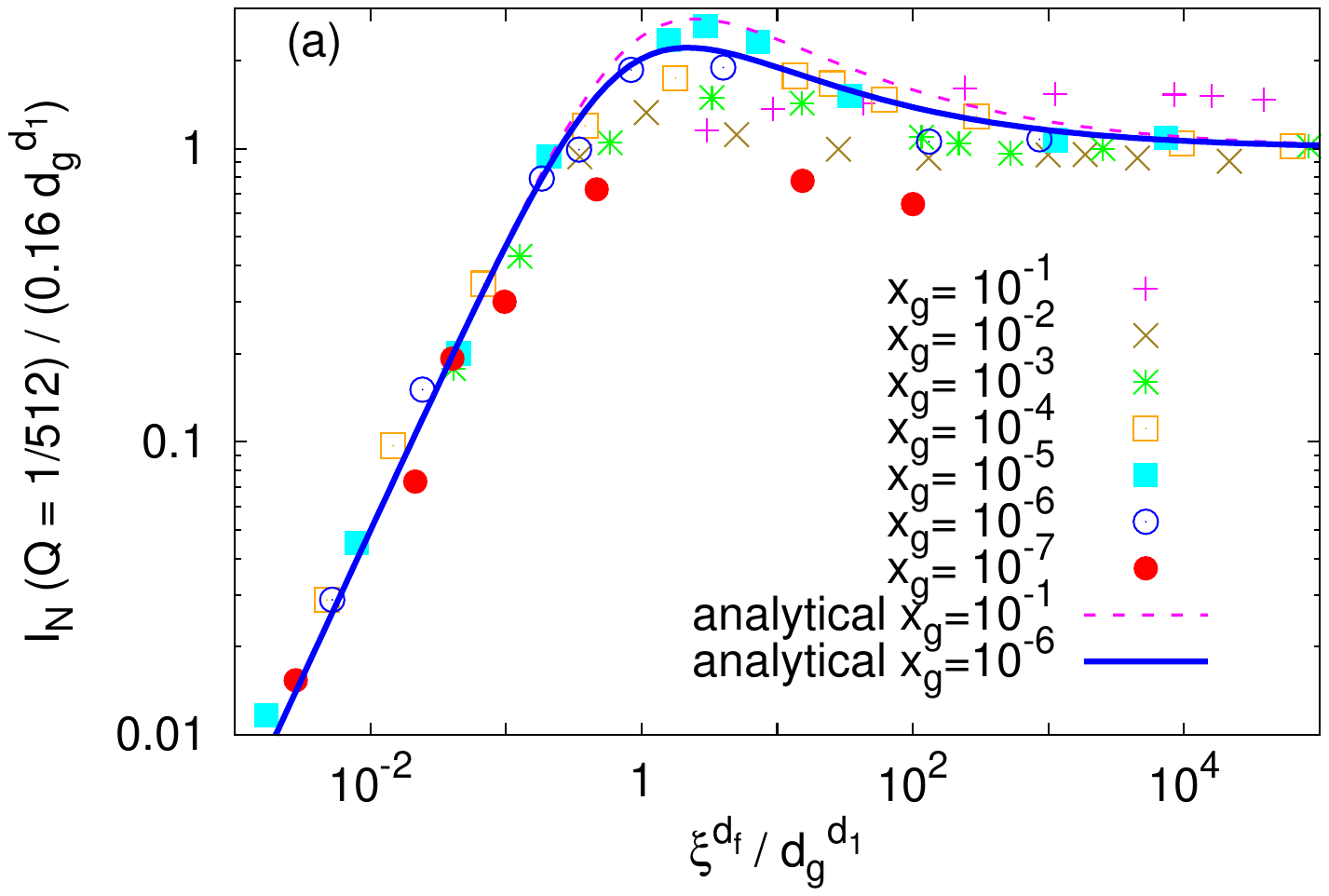}
 \includegraphics[scale=0.3]{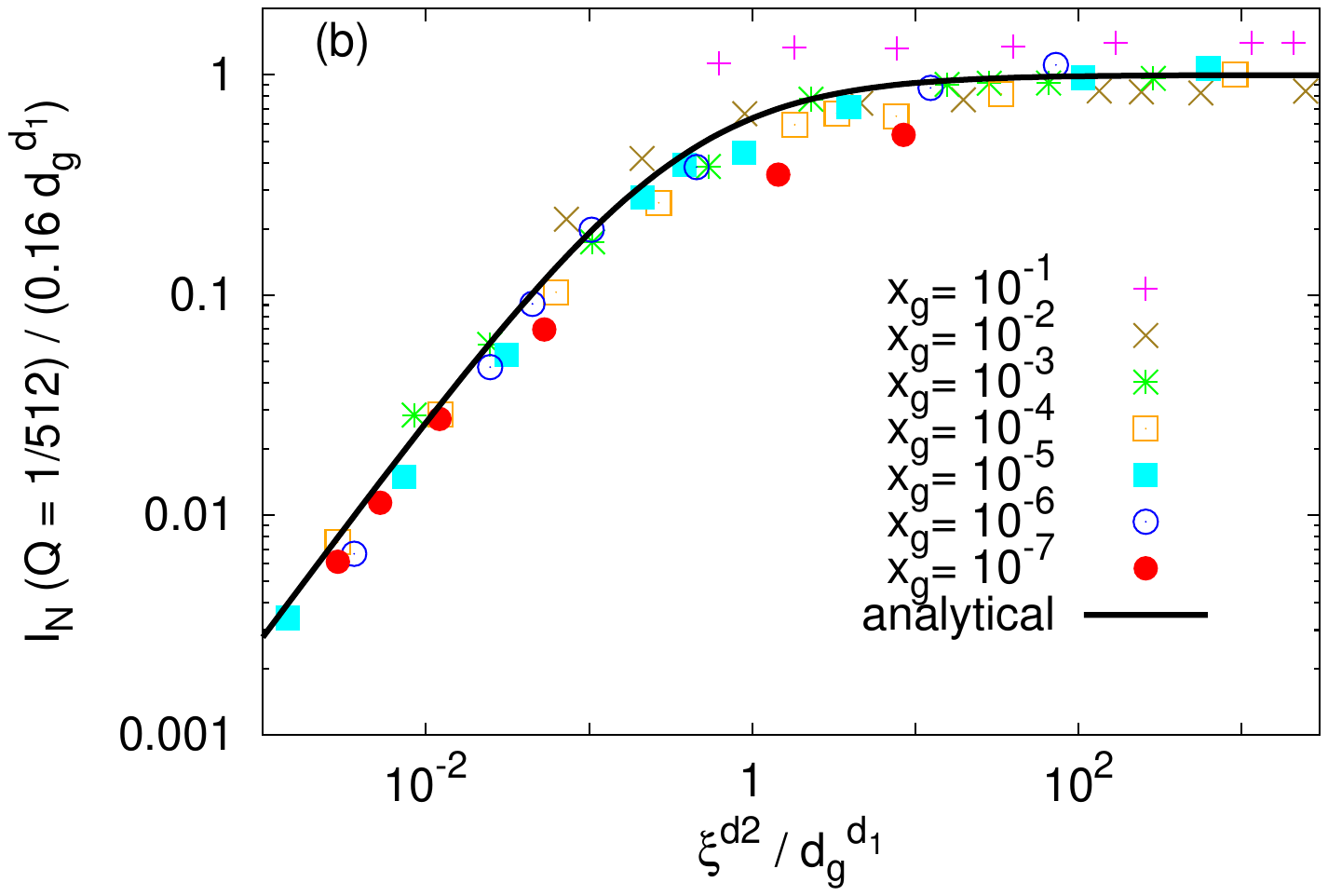}
 \caption{Scaling behavior of the low-$Q$ normalized structure factor, $I_N(Q \to 0)$,  for invasion from germs for all our data (7 germ fractions $\times$ 23 distances from $p_c$); (a) below $p_c$; (b) above  $p_c$. Both figures show $I_N(Q \to 0)$ in scaled coordinates  based on the two characteristic lengths  $\xi$, $d_g$. $I_N(Q \to 0)$ is evaluated at $Q=1/512$, the lowest non-zero $Q$ of the FFT. The exponents $d_1= 2.8$ and $d_2= 2.25$ allow an approximate collapse of all data. The lines correspond to Eq.~\ref{INtheolowQ}.}
\label{fig:scalingSbasQ}
\end{figure}

\medskip
Based on the above analyses, we gather all our simulations results for the low $Q$ limit of the normalized structure factor, $I_N(Q \to 0)$, in the two plots of
 Fig.~\ref{fig:scalingSbasQ}. In these log-log plots, corresponding respectively to below and above $p_c$, the $y$ axis is
  $I_N(Q \to 0)/d_g^{d_1}$. The $x$-axis is $\xi^{d_f}/d_g^{d_1}$ below $p_c$, and  $\xi^{d_2}/d_g^{d_1}$ above.
In these coordinates, the regimes $\xi \ll d_g$ below and above $p_c$ correspond to  straight lines of slope unity,
   and the concentrated regime  $\xi >> d_g$ around $p_c$ to a constant.
    In this representation, all points fall close to unique curves within a typical factor 2.  The largest deviations, observed for $x_g=10^{-7}$, could partly result from the limited sampling in this case (100 germs only).

\medskip
Within the factor 2 above,  the behavior of $I_N(Q \to 0)$ is approximately parametrized by:

\begin{equation}
I_N(Q \to 0, \xi, d_g)= (S_1^{\alpha} (\xi, d_g)+S_2^{\alpha}  (\xi, d_g) )^{1/\alpha}
\label{INtheolowQ}
\end{equation}

with $S_1$, $S_2$, and $\alpha$  given in Table \ref{table:coefffits2}. $S_1$ and $S_2$ are deduced from the asymptotic behaviors in the dilute and concentrated regimes, respectively, except for $S_1$ below $p_c$ which also includes a correcting factor to approximately reproduce the shallow maximum of $I_N(Q \to 0, \xi, d_g)$ for $\xi \approx 2 d_g$ observed in Fig.~\ref{fig:scalingSbasQ}a. $\alpha$ is chosen to match the behavior
in the crossover region  $\xi \simeq d_g$.

\begin{table}[b!]
\begin{center}
\begin{tabular}{c c c c}
\hline 
\hline 
          & $S_1$                      & $S_2$          & $\alpha$ \tabularnewline
\hline 
$p<p_{c}$ & $0.16 \, (1+4d_g/\xi)  \, d_g^{d_1}$ & $0.8  \,\xi^{d_f}$  & $-1$     \tabularnewline
$p>p_{c}$ & $0.16 \, d_g^{d_1}$            & $0.5  \, \xi^{d_2}$ & $-0.8$   \tabularnewline
\hline 
\hline 
\end{tabular}
\end{center}
\caption{ Expressions for the functions $S_1$ and $S_2$ of Eq.~\ref{INtheolowQ}. $d_1=2.8$ and $d_2=2.25$.}
\label{table:coefffits2}
\end{table}

\medskip
Away from $Q \to 0$, the transition of $I_N(Q)$  from the constant low $Q$ limit to the fractal behavior observed at larger $Q$ is found to be reasonably well described for all our results by the simple interpolation. 

\begin{equation}
I_N(Q, \xi, d_g)=[(I_N (Q \to 0, \xi, d_g))^{-1}   + {(0.15 \, Q^{-d_f}})^{-1}]^  {-1}
\label{INtheoQ}
\end{equation}

The $Q$ dependence predicted by this equation is shown by continuous lines in Figs.\ref{fig:SO-Q-germs-belowpc} and~\ref{fig:SdeQaprespc}.

\medskip
By allowing to estimate the effect of inter-cluster correlations for invasion from bulk germs, Eq.~\ref{INtheoQ} is a central result of our simulations. 
In order to obtain the absolute structure factor per unit volume 
$I(Q)= p_{\rm{eff}} (1-p_{\rm{eff}}) I_N(Q)$, which is the quantity directly measured in a scattering experiment, Eq.~\ref{INtheoQ} has to be complemented by expression(s) for $p_{\rm{eff}}( p, x_g)$. We will use the following equations:

\begin{equation}  \label{pefftheoavantaprespc}
p_{\rm{eff}} ( \xi, d_g) = \begin{cases}
x_g \, d_{g}^{d_f} \, [0.41^{\beta} + (0.68\,  \xi ^{(3-\tau) d_f} / d_{g}^{d_f})^{\beta} ] ^{1/\beta}  & p<p_c \\
x_g \, d_{g}^{d_f} \, [0.41^{\gamma}+ (0.55\,  (\xi/d_g)^{d_f-3})^{\gamma})]^{1/\gamma}            & p>p_c
\end{cases}
\end{equation}

with $\beta=-0.9$ and $\gamma=3$.

 \medskip
For $\xi \gg d_g$ or $\xi \ll d_g$, these equations reduce to the limiting behaviors found above: $p_{\rm{eff}} \approx x_{g} d_{g}^{d_f}$ close to $p_c$,
     $p_{\rm{eff}} \approx x_g  \, \xi^{(3-\tau) d_f}$ below $p_c$ in the dilute regime, and
      $\xi^{(d_f-3)}$ above $p_c$ in the concentrated regime. Values of $\beta$ and $\gamma$ are chosen to reproduce the simulation results in the intermediate region $\xi \sim d_g$ (see the continuous lines in Fig.~\ref{fig:scalingpeff}). 

\begin{figure}[ht]
 \includegraphics[scale=0.3]{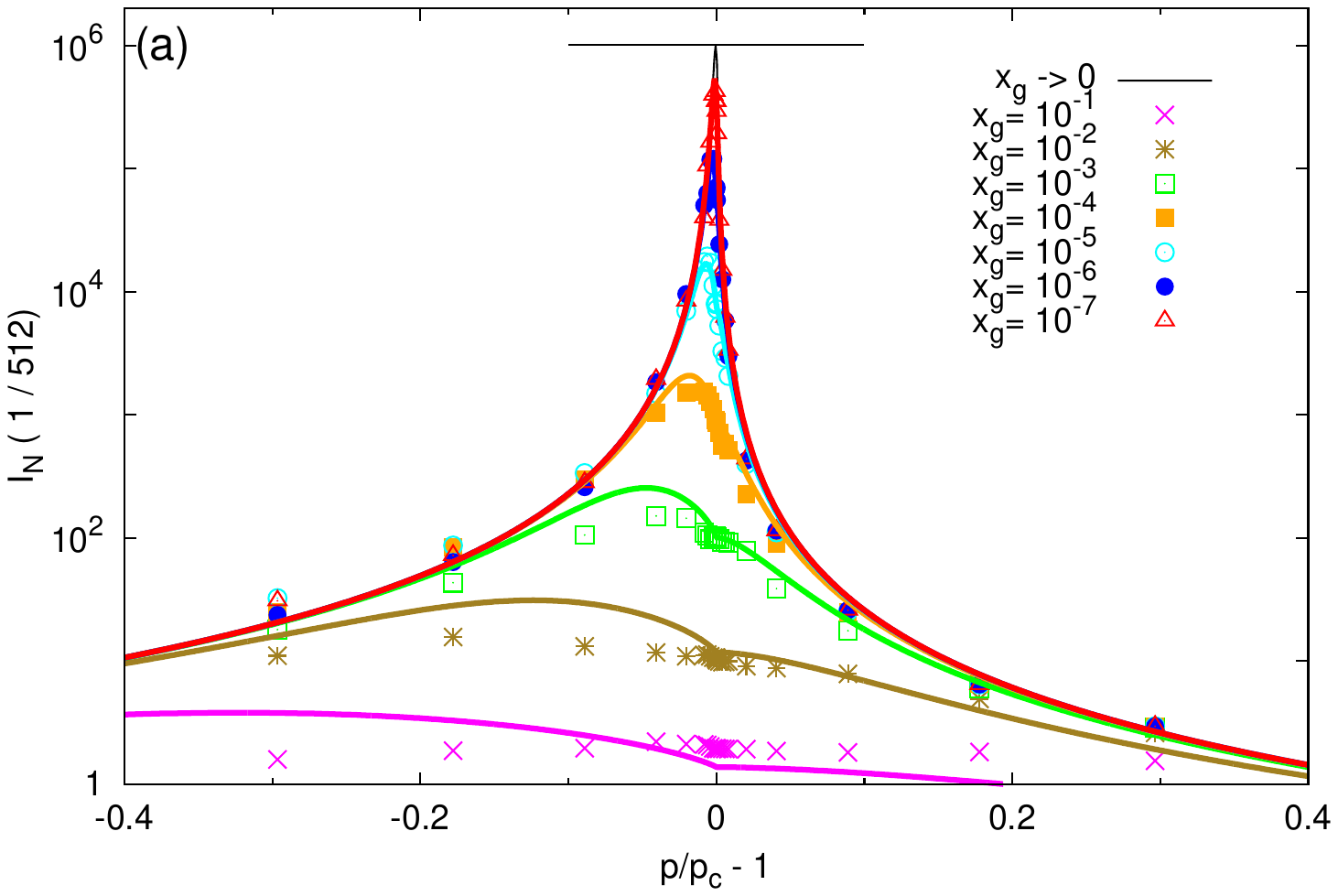}
\includegraphics[scale=0.3]{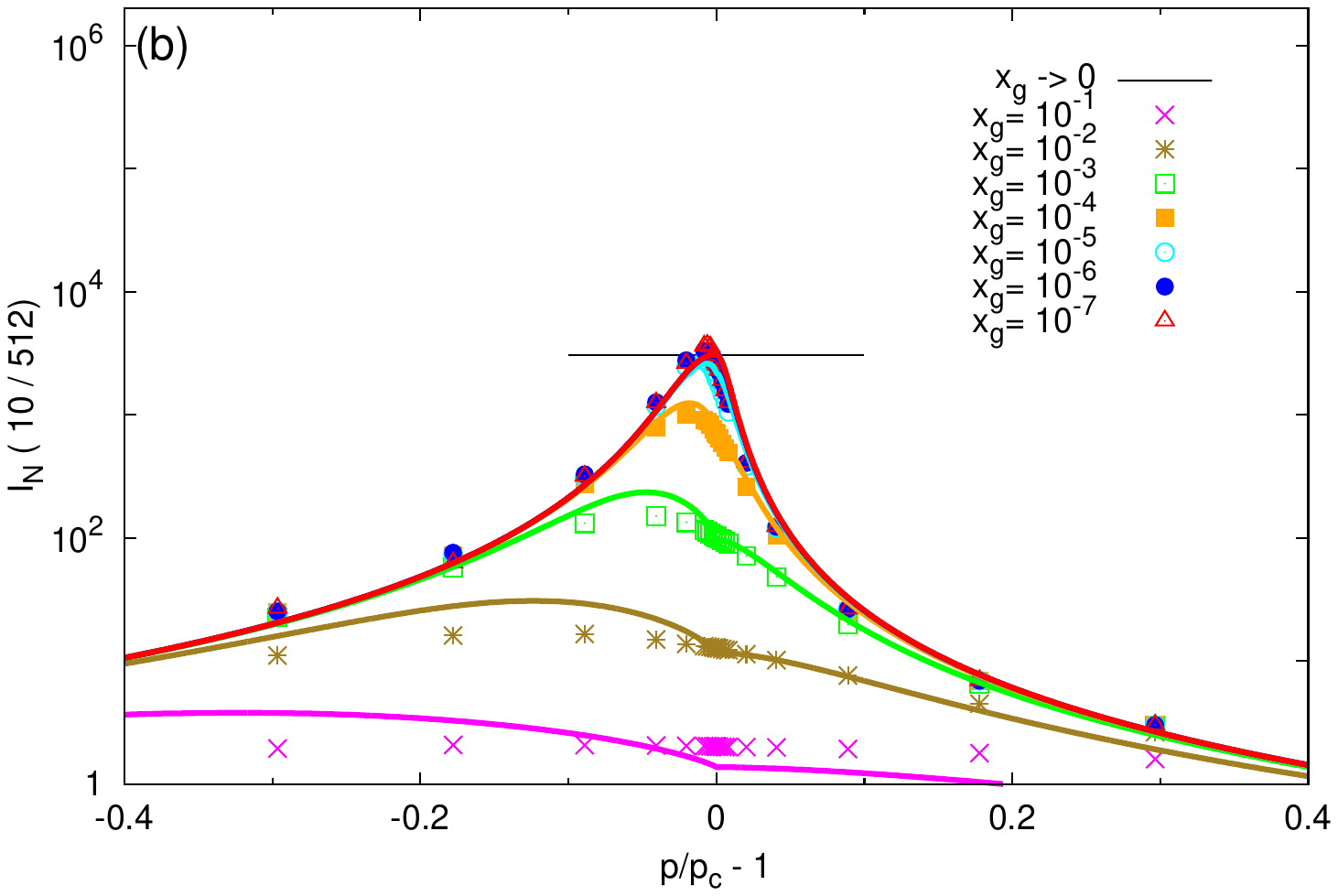}
\includegraphics[scale=0.3]{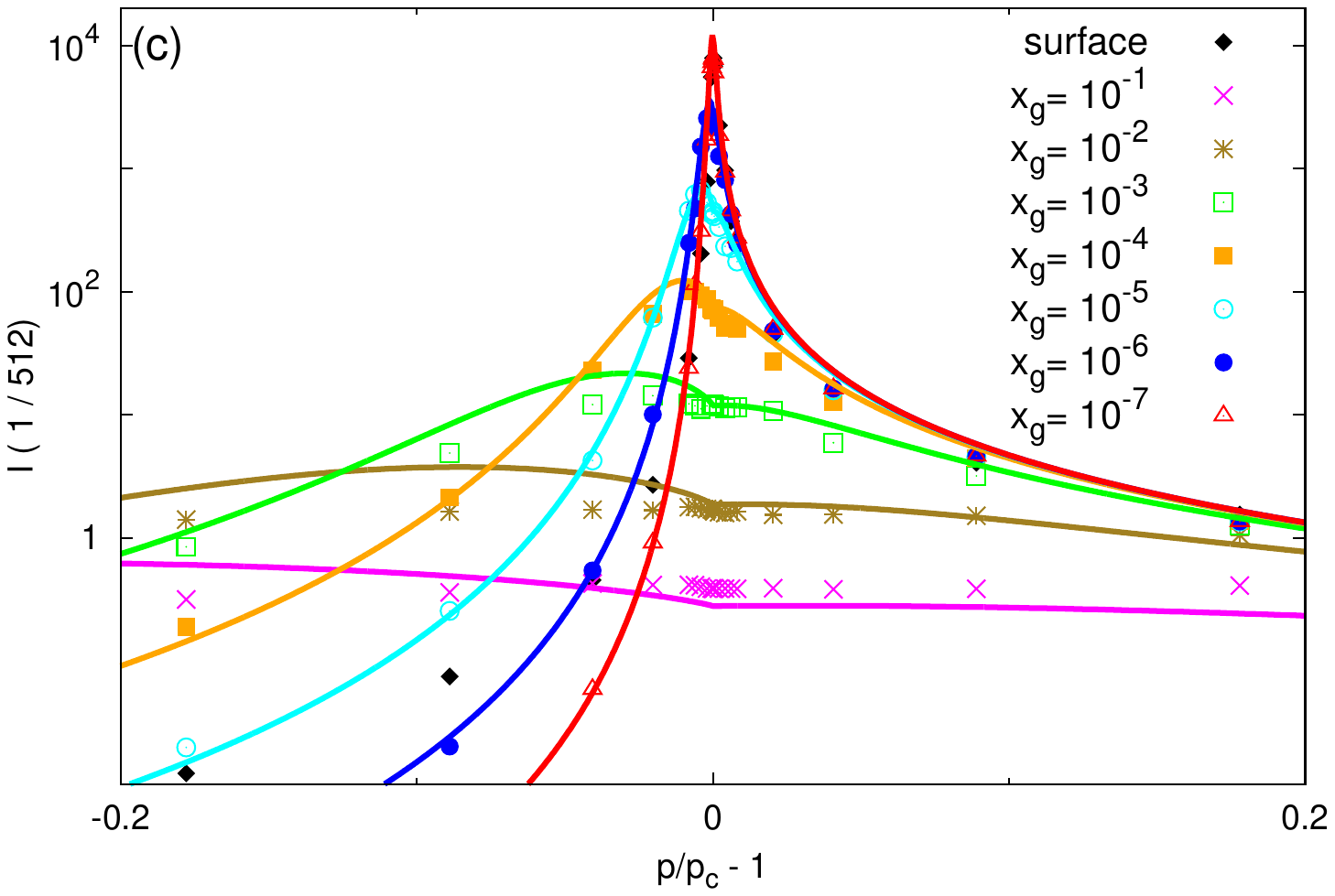}
\includegraphics[scale=0.3]{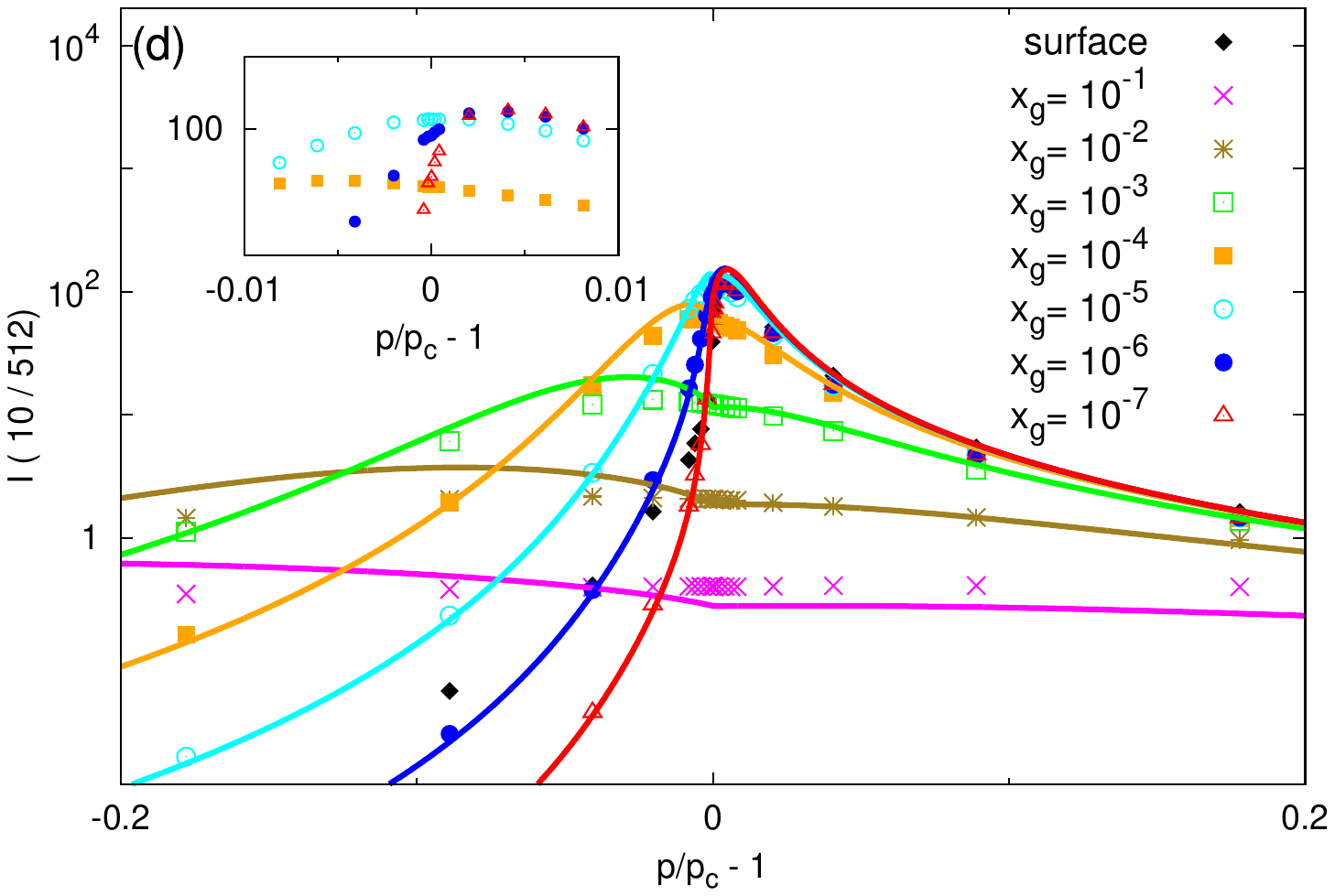}
 \caption{Dependence of the normalized (a,b) and absolute (c,d) structure factors $I_N (Q, p, d_g)$ and $I (Q, p, d_g)$  around the percolation threshold for (a,c): $Q=2/L\approx 0.002$, and (b,d): $Q=20/L\approx 0.02$. Symbols correspond to the results of simulations for germ fractions from $10^{-7}$ to  $10^{-1}$.
   Continuous lines give the predictions of
   Eqs.~\ref{INtheolowQ},~\ref{INtheoQ}~and~\ref{pefftheoavantaprespc}
   for the same fractions, plus a vanishingly small fraction (corresponding to the limit $d_g=\infty$ in Eq.~\ref{INtheolowQ}). The horizontal bars in figures (a) and (c) correspond to the fractal limit of $I_N(Q)$, $0.15 \, Q^{-d_f}$, the maximal degree of coherence at wavevector $Q$.  Figures (c) and (d) also
show the absolute structure factor $I(Q, p)$ for invasion from surfaces as filled diamonds. The insert in figure (d) is a zoom on the peak region. 
   }
\label{fig:SQnorm}
\end{figure}

\medskip
To summarize this section, eqs.~\ref{INtheolowQ},~\ref{INtheoQ} and~\ref{pefftheoavantaprespc} properly describe the normalized and absolute structure factors for invasion percolation from bulk germs. While the choice of values for the parameters ($d_1$, $d_2$, $\alpha$, $\beta$, $\gamma)$ is somewhat arbitrary, it reasonably reproduces, for the whole set of ($p$,$x_g$) parameters used in our simulations, the effects of the cluster size distribution and of the inter-cluster correlations.
 This agreement will be again illustrated in section~\S\ref{sec:Discussion}, which discusses the $p$ dependence of $I_N(Q)$ and $I(Q)$, depending on the germ fraction and  the scattering wave vector $Q$. Moreover, we will show in section~\S\ref{sec:Evaporation} that eqs.~\ref{INtheolowQ},~\ref{INtheoQ} and~\ref{pefftheoavantaprespc} can be successfully used to predict the scattering properties for values of ($p$,$x_g$) different from those simulated above.

\subsection{Discussion of the scattering signal for invasion from bulk germs at a fixed fraction}
\label{sec:Discussion}
 
In this section, we discuss the $p$ dependence, for the different germ fractions, of the normalized and absolute structure factors obtained from our simulations for two values of $Q$,  $Q=2/L\approx 0.002$, the lowest non-zero $Q$ value, and $Q=20/L\approx 0.02$, 10 times larger. This dependence is shown in Figs.~\ref{fig:SQnorm}a and \ref{fig:SQnorm}c, and \ref{fig:SQnorm}b and \ref{fig:SQnorm}d,  respectively. While these figures contain no new data with respect to the contents of Figs.\ref{fig:scalingpeff}  and~\ref{fig:SQnorm}, the representation in function of $p$ is more adapted to a discussion of the effect of percolation on scattering, and to a comparison to experiments. In these figures, symbols correspond to the results of simulations, while continuous lines correspond to the predictions of Eqs.~\ref{INtheolowQ},~\ref{INtheoQ}~and~\ref{pefftheoavantaprespc}, using for the correlation length the following expressions
 
\begin{equation}\label{eq:fitxi2bashaut}
  \xi= \begin{cases}
    1.25 \, (1-p/p_c)^{-0.876}   &   p<p_c \\
    0.681  \, (p/p_c-1)^{-0.929}  &   p>p_c
  \end{cases}
\end{equation}

which properly reproduce our simulations results for the four configurations studied  and $\xi /L <  0.14 $ (see Appendix A). 

\medskip
Apart from a slope discontinuity at $p_c$, and discrepancies with simulations for fractions of  $10^{-1}$ and  $10^{-2}$, for which the scattering is anyway small,  the overall agreement is good. 
  In particular, Eqs.~\ref{INtheoQ} and ~\ref{pefftheoavantaprespc} 
  well account for the evolution of the peaks of $I_N(Q,p)$ and $I(Q,p)$
    both in height, position with respect to $p_c$, and dependence upon $Q$.
    In the following, we discuss the salient points of these figures in the perspective of the experimental detection of percolation through scattering measurements.

\medskip
First, a clear scattering peak at percolation is only observed for a small enough germ fraction. 
If the germ fraction is  too large ($\geq 10^{-2}$), the distance between germs is so small that the condition $\xi \sim d_g$ corresponds to $p$ well below $p_c$. As a consequence, the peak occurs well before the percolation threshold and the maximal gain of coherence is modest.

 \medskip
Second, the peak of the normalized structure factor becomes sharper and higher as the distance between germs increases,
 and the fractal correlations can correlatively develop up to a larger scale.
   Looking in more detail,  Figs.\ref{fig:SQnorm}a and \ref{fig:SQnorm}b show that, for the considered $Q$'s,
    $I_N(Q,p)$ barely depends on $Q$ except close to $p_c$. 
    This behavior corresponds to the plateau of Fig.~\ref{fig:SO-Q-germs-belowpc}a at  $Q.\xi <1$.
     In this regime, as long as $d_g>\xi$, corresponding to  $x_g < \xi^{-3}$, $I_N(Q,p)  \approx \xi^{d_f}$ does not depend on the germ fraction either. 
      In contrast, close to $p_c$, $\xi$ is large enough for the coherence to be limited by either $d_g$ or $1/Q$. $I_N(Q,p)$ then increases when the germ fraction or $Q$ decrease. In particular, for a low enough germ fraction (such that both $Q.\xi $ and $Q.d_g$ are larger than unity), $I_N(Q,p)$ becomes limited by the fractal behavior $Q^{-d_f}$.

 \medskip
Third, Figs.\ref{fig:SQnorm}c and \ref{fig:SQnorm}d show that the absolute structure factor $I(Q,p)$ presents similar trends.
In particular, it strongly depends on $Q$ only close to $p_c$ and for germ fractions smaller than $10^{-5}$. 
It is for these conditions that the fractal signature of percolation is the clearest.
This is specially true for the absolute structure factor, due to the fast increase of the 
the $p_{\rm{eff}}$ term when $p$ increases up to $p_c$.

 \begin{figure}[ht!]
\includegraphics[scale=0.5]{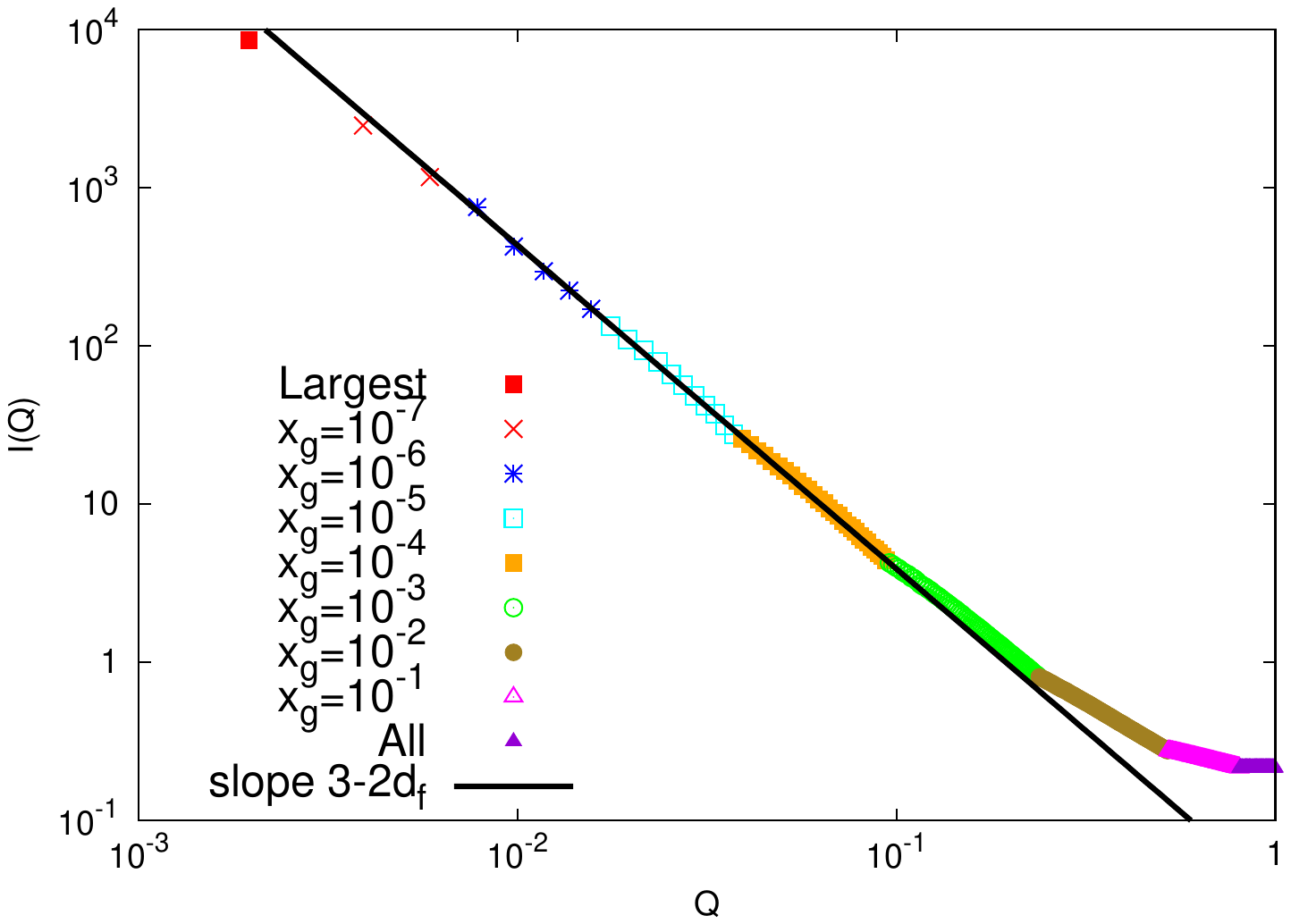}
\caption{$Q$ dependence of the maximum value of the absolute structure factor among all our
  results for different selections (bulk germs, largest cluster, and all clusters) and for all $p \leq p_c$. In the rare instances where the maximum is reached for $p<p_c$, its value differs from the value at $p_c$ by less than 0.1\%. The symbols indicate which selection maximizes the structure factor. 
  The $Q^{3-2d_f}$ dependence, and the fact that the maximum is obtained for larger fractions at larger $Q$ are in agreement with the arguments given in
the text.}
\label{fig:maximumabsoluteAll}
\end{figure}

\medskip
The above results allow to discuss a practical question which sets the ability to detect percolation through scattering measurements.
For a given $Q$ (i.e.  scattering angle), what is the maximal absolute signal, and at which germ fraction does it occur ?
The existence of  an optimal fraction follows from the competition between the opposite dependences of $p_{\rm{eff}}$ and $I_N(Q \to 0,p)$ on the germ fraction.
  At small germ fractions, such that $\xi < d_g$, $I_N(Q \to 0)$ is  a constant of order $\xi^{d_f}$ and $I(Q,p)$ thus increases with $p_{\rm{eff}}\propto x_g$.
At larger fractions, corresponding to $d_g < \xi$, $I_N(Q, d_g) \sim d_g^{d_f}$ drops faster than $p_{\rm{eff}}$ increases, and $I(Q\to 0)$ decreases.
Combining $p_{\rm{eff}} \propto d_g^{d_f-3}$ and  $I_N(Q \to 0) \propto d_g^{d_1}$, we find that 
$I(Q\to 0)\propto d_g^{d_f+d_1-3}$, which decreases approximately as 
$d_g^{2 d_f-3} \sim x_g^{-2/3}$ as $x_g$ increases. 
Neglecting the difference between $d_1$ and $d_f$,
 the maximal value of $I(Q\to 0 ,p,d_g)$ is of order $d_g^{2 d_f-3}$, obtained for $d_g \simeq \xi$, which increases with $\xi$.
 However, for the $Q \to 0$ limit to hold, $\xi $ should remain smaller than $Q^{-1}$.
  For larger values, $I_N(Q) \sim Q^{-d_f}$, and $I(Q) \propto p_{\rm{eff}}$ which decreases when $d_g \sim \xi $ increases. 
Hence, our analysis predicts that the maximal absolute signal at a given $Q$
 is of order $Q^{3-2 d_f} \simeq  Q^{-2}$,
 obtained for $d_g \simeq 1/Q$ and a distance to $p_c$ such that $\xi \simeq d_g$,
 corresponding to the situation where the selected clusters are close to interconnect. 
 
 \medskip
 In order to directly check these conclusions, we have extracted from our results, for each value of $Q$, the configuration (among the selections of the largest cluster, all clusters, or bulk germs at a fixed fraction) maximizing the absolute structure factor.
We report in Fig.~\ref{fig:maximumabsoluteAll} the corresponding value  of $I(Q)$ as a function of $Q$, the symbols coding for the configuration maximizing the signal.
Except for the lowest $Q$ (where the largest cluster gives the maximal structure factor) and the largest $Q$ (where the maximum is obtained by selecting all clusters), the maximum of the absolute structure factor corresponds to the case of invasion from bulk germs.
The predicted $Q^{3-2d_f}$ dependence, and the fact that the maximum is obtained for larger germs fractions at larger $Q$ are accurately 
confirmed by Fig.~\ref{fig:maximumabsoluteAll}. 
 
\medskip
The  $Q^{3-2d_f}$ dependence of the maximal absolute structure factor quantifies 
the physical expectation that percolation is more easily detected at small $Q$ values (small scattering angles).  
Moreover, a second reason for using small scattering angles emerges from the inset of Fig.~\ref{fig:SQnorm}b.
Zooming on the region close to $p_c$ for $Q=20/L$ reveals that, for small germ fractions,  
the maximum of $I(Q ,p,d_g)$ occurs beyond percolation, 
the growth of $p_{\rm{eff}}$ offsetting the decrease of $I_N(Q ,p,d_g)$.
This behavior is observed not only for our simulations, but also for the predicted curves, 
showing that it is not due to finite size or sampling errors.
This implies that precisely measuring the percolation threshold through the position of the maximal scattering signal 
requires not only the germ fraction to be small, but also $Q$.

  \medskip
To summarize, percolation invasion from bulk germs at a fixed small fraction has a clear scattering signature at small scattering angles. 
This result is qualitatively obvious. Indeed, in such conditions, clusters can grow up to a large size before the concentrated regime is reached.  
What our study allows is to quantify these conditions. As an example, according to Fig.~\ref{fig:maximumabsoluteAll} or Fig.~\ref{fig:SQnorm}b, for $Q=0.02$, corresponding to probing a scale 50 times the mesh size, the scattering signal increases by a factor 2500 with respect to its incoherent value as long as the germ fraction remains smaller than 10$^{-5}$.
This extended dynamic range allows the detection of percolation, and a test of the fractal structure of the clusters over a large range of scales, even in the case where the incoherent scattering is small and/or masked by some background. 
 Furthermore, as shown by the inset of Fig.~\ref{fig:SQnorm}d, even though the absolute signal peaks at $p<p_c$, the relative deviation of $p$ from $p_c$  is less than about 1\%, for $x_g \leq 10^{-4}$. Hence, for such small germ fractions, the percolation transition nearly corresponds to the peak of scattering. Note however that, as illustrated in the next section, this conclusion fails if the germ fraction is too large.

 \subsection{Application to evaporation in a 3D connected disordered porous material}
\label{sec:Evaporation}
In this section, we discuss evaporation from a 3D connected porous material in the regime of percolation invasion from bulk germs created by thermally activated cavitation. In this regime,  both $x_g$ and $p$ depend on pressure and temperature. Hence, in contrast to the previous section,$x_g$ the germs fraction is no longer constant, but increases with $p$. In order to evaluate the resulting structure factor, a first approach is to rely, as above, on specific simulations, using a physical model to compute the input parameters $x_g$ and $p$. Given these parameters, an alternative, though approximate, approach, is to apply Eqs.~\ref{INtheolowQ},~\ref{INtheoQ}~and~\ref{pefftheoavantaprespc}. In the following, we first use the simulation approach to determine and to discuss the temperature dependence of the scattering signal. We then show that the second approach properly reproduces, at nearly no numerical cost, the main features of this temperature dependence, thereby demonstrating its interest.

 \medskip
As in Ref.\citenum{Bonnet2019b}, we model the porous material  by pores of randomly distributed radius $R$ occupying the sites of a cubic lattice.
At given temperature $T$ and pressure $P$,  $p$ and $x_g$  are respectively given by the fractions of pores such that  $P_{\rm{eq}}(R,T) >P $ and $P_{\rm{cav}}(R,T) >P$, , where $P_{\rm{eq}}$ and $P_{\rm{cav}}$ are defined in the introduction. $P_{\rm{eq}}(R,T)$ and $P_{\rm{cav}}(R,T)$ both increase with $R$,  so that $p(P,T)$ and $x_g(P,T)$ are respectively the fraction of pores of radius $R>R_{\rm{eq}}(P,T)$ and $R>R_{\rm{cav}}(P,T)$, where $R_{\rm{eq}}$ is such that $P_{\rm{eq}}(R_{\rm{eq}},T)=P$ and similarly  for $R_{\rm{cav}}$.
For a given pore radius distribution, $p(P,T)$ and $x_g(P,T)$ can be determined from the radius and temperature dependences of $P_{\rm{eq}}(R,T)$ and $P_{\rm{cav}}(R,T)$ computed using a model for evaporation in a single pore, such as that developed in Ref.\citenum{Bonnet2019a}. This allows to obtain the -temperature dependent- percolation pressure $P_{\rm{perco}}$ such that $p(P_{\rm{perco}},T)=p_c$ and a crossover temperature $T^*$ such that $P_{\rm{cav}}^{\infty}(T^*)=P_{\rm{cav}}(R \to \infty,T^*)$ is smaller than $P_{\rm{perco}} (T^*)$. Below $T^*$, no bulk germ is present for $P \geq P_{\rm{perco}}$, so that invasion percolation from bulk germs only takes place above $T^*$.

\begin{figure}[ht!]
\includegraphics[scale=0.3]{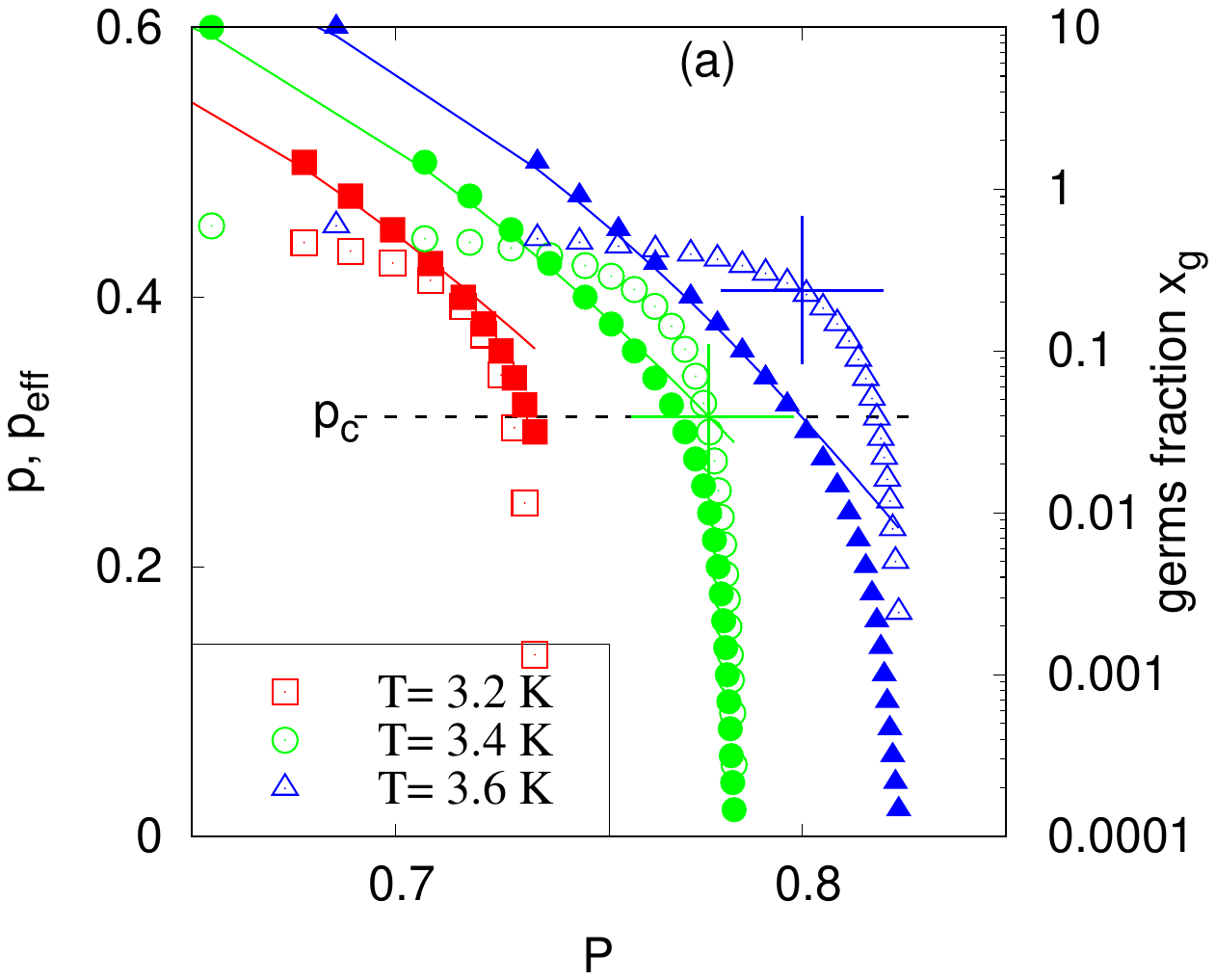}
\includegraphics[scale=0.3]{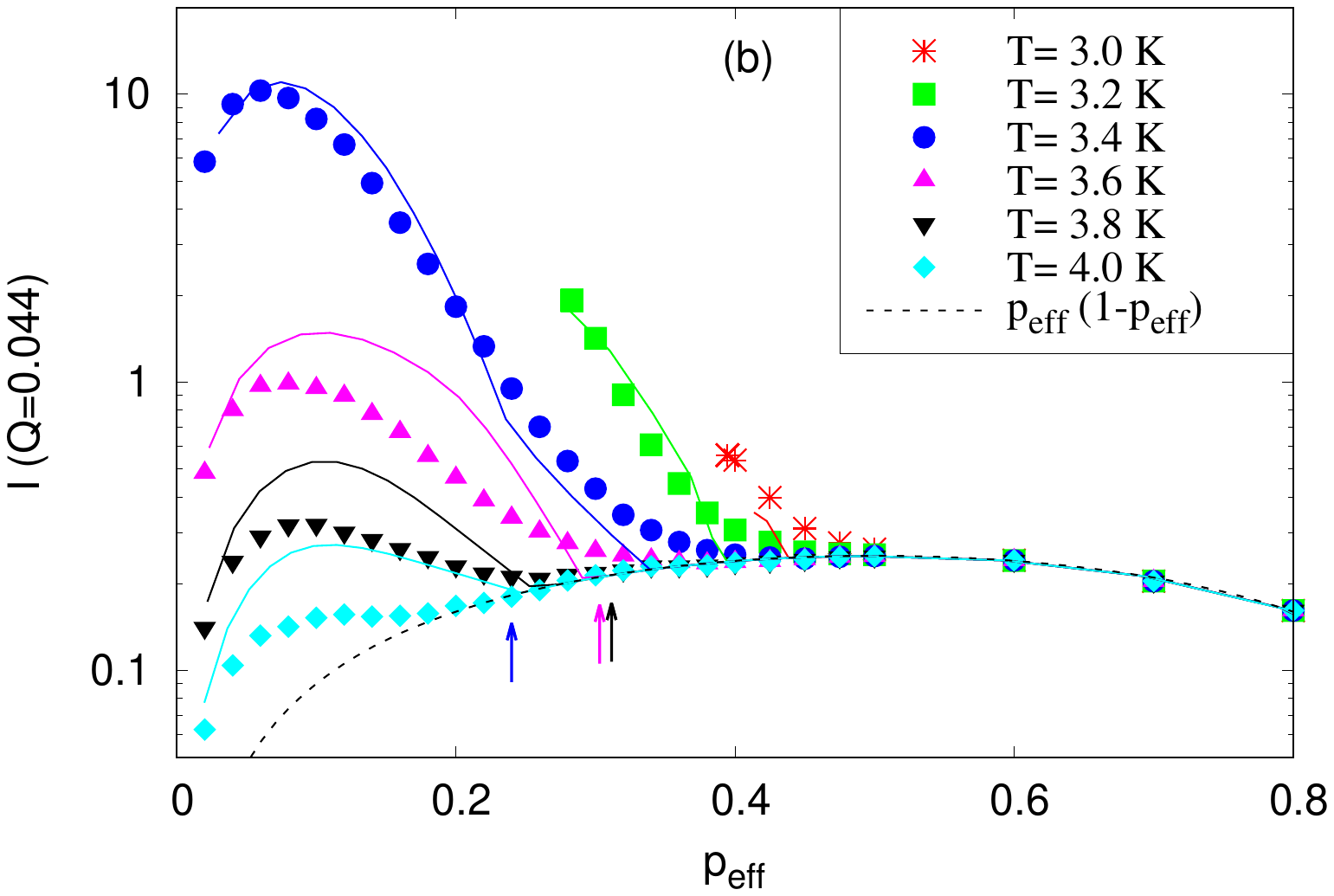}
\caption{Theoretical optical signature of percolation invasion from bulk germs for evaporation of helium in a porous material;(a): Dependence of the parameters $p$ (continuous line) and $x_g$ (open symbols) on the pressure $P$ at three temperatures for $P< P_{\rm{cav}}^{\infty}$ (i.e.   a non zero germ fraction) as analytically computed for a gaussian distribution of pore radius (see text). Both $p$ and $x_g$  increase when the pressure decreases or the temperature increases.
 For $T > T^* = $ 3.35 K, crosses indicate the value of the germ fraction at the percolation threshold.
Closed symbols give $p_{\rm{eff}}(P)$ as obtained from simulations on a $L$=512 pore lattice generated using the same radius distribution. (b) Absolute structure factor obtained  from these simulations for $Q=0.044$ as a function of $p_{\rm{eff}}$ at different temperatures (symbols). At large  $p_{\rm{eff}}$, $I_N(Q)$ approaches the incoherent limit 
   $p_{\rm{eff}}
   (1-p_{\rm{eff}})$, given by the dotted line.
     The value of $p_{\rm{eff}}$  at percolation is indicated by arrows. It coincides with $p_c$ above 3.8~K. Continuous lines
   give the structure factor predicted from $p$ and $x_g$ using  
   Eqs.~\ref{INtheolowQ},~\ref{INtheoQ}~and~\ref{pefftheoavantaprespc}.
   These predictions correctly reproduce the main trends of the simulation results. }
\label{fig:compareLangmuir}
\end{figure}

\medskip
Figure~\ref{fig:compareLangmuir}a shows the pressure, or, more precisely, the fugacity \footnote{As in Ref.\citenum{Bonnet2019b}, we denote by $P$ the fugacity of the vapor, which, in the limit of the perfect gas approximation, reduces to the ratio of the vapor pressure to the saturated vapor pressure.} dependence of $p$ and $x_g$  for the  physical conditions studied in Ref.\citenum{Bonnet2019b}, namely liquid helium in the range 3-4 K and pores radii $R$ distributed according to a gaussian of mean 3~nm and width 1.31~nm.
 The three temperatures shown in
 Fig.~\ref{fig:compareLangmuir}a are below (3.2~K), above (3.6~K), or close (3.4~K) to the crossover temperature $T^*$ (3.35~K for the above parameters~\cite{Bonnet2019b}). At a given temperature, the cavitated pores act as vapor germs and the pores effectively emptied are those belonging to the clusters containing these germs. The germ fraction at percolation is zero below $T^*$ and rapidly increases with temperature above: as shown by Fig.~\ref{fig:compareLangmuir}a,$x_g(p_c)$ is  larger than $10^{-2}$ for $T \geq$ 3.4 K.
 
\medskip
In the simulation approach,  we evaluate $p_{\rm{eff}}$ and  $I(Q)$ using a single realization of a $L=512$ lattice where the pore radius at each site is drawn according to the gaussian distribution above.
 At each pressure $P$, we select the clusters containing the germs and compute $p_{\rm{eff}}$ and  $I(Q)$ as described in \S\ref{sec:Selection} and \S\ref{sec:StructureFactorSelected}  \footnote{The scattering intensities computed in Ref. [8] or shown in Fig.1(b) include the scattering by the porous matrix. This explains why, in contrast to $I(Q)$,  they do not tend to 0 as $p_{\rm{eff}}$ tends to 0 or 1.}.  Figures~\ref{fig:compareLangmuir}a and b respectively show the results for $p_{\rm{eff}}(T)$ as a function of $P$ and $I(Q,T)$ 
as a function of $p_{\rm{eff}}$ for different temperatures and $Q=0.044$ (this value corresponding to the scattering angle in Fig.\ref{fig:Experiments}b).
 
 \medskip
 Above $T^*$, a peak of the structure factor develops during evaporation. In agreement with the experimental behavior shown in Fig.~\ref{fig:Experiments}b, its amplitude 
 increases when the temperature decreases. 
In contrast to the situation shown in Fig.~\ref{fig:SQnorm}, the peak occurs well before percolation. 
 This difference follows from the fast increase of the germ fraction below $P_{\rm{cav}}^{\infty}$ (see Fig.~\ref{fig:compareLangmuir}a). 
Except in a narrow interval above $T^*$, $d_g \sim\xi$ occurs for a small value of $d_g$, hence $\xi$,  accounting for the peak occuring well below $p_c$.

 \medskip
   Below $T^*$, cavitation in the largest pores takes place below $P_{\rm{perco}}$. Because we 
 considered periodic boundary conditions, thus excluding invasion from surfaces,
   $p_{\rm{eff}}$
   and the structure factor abruptly jump from zero to finite values at  $P_{\rm{cav}}^{\infty}$.  At 
   lower pressures,  $p_{\rm{eff}}$  increases, but the decrease of $d_g$ makes $I(Q)$ to decrease, accounting for the decrease of $I(Q)$ with increasing $p_{\rm{eff}}$ 
   observed in Fig.~\ref{fig:compareLangmuir}b for $T=3.0$ and $3.2$~K.
    Finally, at all 
   temperatures, $I(Q)$ at large enough values of $p_{\rm{eff}}$ is given by the incoherent limit  
   $p_{\rm{eff}} (1- p_{\rm{eff}})$. 
   This results from the fact that, below a temperature dependent critical radius, $P_{\rm{cav}}(R,T)$ coincides with  $P_{\rm{eq}}(R,T)$ \cite{Bonnet2019a}.
    Below this commun pressure, all potential sites are germs, so that all the clusters are selected and the structure factor coincides with that for a random distribution with  $p_{\rm{eff}}=p$.
     This high $p_{\rm{eff}}$ regime is a direct illustration of the key role of the selection process in the obtention of a coherent enhancement of $I(Q)$.

\medskip
The predictions based on 
  Eqs.~\ref{INtheolowQ},~\ref{INtheoQ}~and~\ref{pefftheoavantaprespc}
are shown by continuous lines in Fig.~\ref{fig:compareLangmuir}b \footnote {Because the comparison explores ranges of $p$ and $x_g$ not studied in this paper (far below $p_c$ and $x_g >$10\%),
   the expression of $p_{\rm{eff}}$ could exceed $p$, which is not possible. In this case, we set $p_{\rm{eff}}$ to $p$ to compute  $I (Q)$.}. Comparison to the closed symbols shows that this second approach correctly accounts for the results of the first one. Specifically, the temperature dependence of the  position and height
     of the peak of $I(Q)$, as well as of the range of $p_{\rm{eff}}$'s
     over which the coherent signal vanishes, are well reproduced.
 Considering that the expressions for $I (Q)$ are only precise within a factor 2,
  and that Fig.~\ref{fig:compareLangmuir}b explores ranges of $p$ and
  $x_g$ not included in the determination of
  Eqs.~\ref{INtheolowQ} and ~\ref{pefftheoavantaprespc}
  (far below $p_c$ and $x_g >$10\%), this agreement is quite satisfying.
 
 \medskip
  This comparison demonstrates that the approximate expressions presented for $I_N(Q)$ and $p_{\rm{eff}}$ can be efficiently used to predict $I(Q)$ for radius distributions or fluid properties different from those simulated here, without having to resort on specific and numerically expensive simulations. They thus provide a tool to analyze the role of percolation as an evaporation process in future scattering experiments.
 
\section{Surface germs}
\label{sec:SurfaceGerms}

\begin{figure}[ht]
  \center
  \includegraphics[scale=0.32]{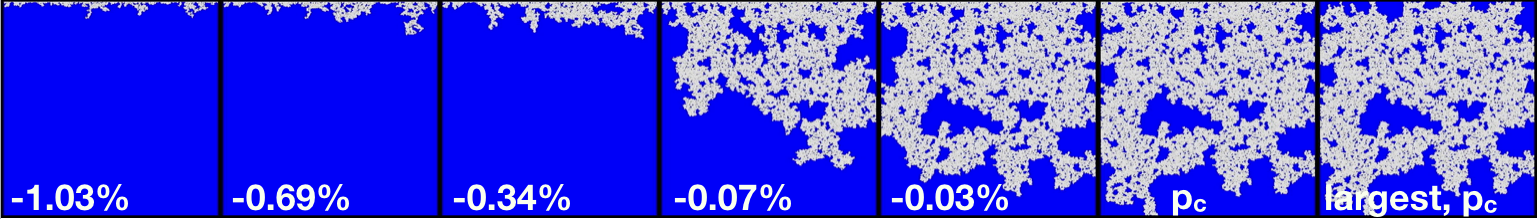}
  \caption{Invasion from surfaces: the selected clusters are those touching the top side of the same 2D lattice as in Fig.~\ref{fig:pictures_clusters}. Periodic boundaries conditions are used only along the direction parallel to this boundary. Labels indicate the relative distance to the percolation threshold, which varies from -1\% to 0\%. For comparison, the last image shows the largest cluster right at $p_c$.
    }
  \label{fig:pictureboundaryclusters}
\end{figure}

In this last section, we discuss the structure factor for invasion from the boundaries of the
 sample when no bulk germs are present. For evaporation from porous materials,
  this situation occurs when the temperature is too low for thermally activated cavitation
   to take place and to create bulk germs. In this case, only clusters connected to the
    surfaces of the sample in contact with the outer vapor can empty.
     The result of the selection process for increasing $p$ values is illustrated
      in Fig.~\ref{fig:pictureboundaryclusters} in the 2D case,
       when only clusters contacting the top side of the square are considered. 
  Comparison of this figure to the first row of Fig.~\ref{fig:pictures_clusters} shows that,
   in contrast to the case of bulk germs, one can never identify a dilute regime. 
   Close to the surface, the density of germs is always large.
    In particular, in the first layer parallel to the surface, all potential sites 
act as germs. Their typical separation,  $p^{-1/2}$ (at 3D),
       is thus of order unity for $p>0.1$. 
Due to this large germ density, it is not obvious that percolation should have a clear-cut signature in terms of light scattering. We show below that this is nevertheless the case.

\medskip
Due to the anisotropic character of the distribution of selected sites in the present case,
   obtaining the structure factor requires to modify its computation with
    respect to the case of bulk germs. Using the same averaging scheme over the
     different orientations of $\textbf{Q}$ at fixed $Q$, and selecting all clusters
      connected to the six sample's surfaces would  result in
       a spurious $Q$ dependence at small $Q$. 
 Indeed, in  the directions ($\pm Q$, 0, 0), the large number of sites along the surface (0, $y$, $z$) 
 give a strong coherent contribution, and similarly for the other
    surfaces. Since, at three dimensions,  the number of $\textbf{Q}$ vectors
of modulus $Q$ scales as $Q^2$, the singular directions give an
  extracontribution diverging as $1/Q^2$ at small $Q$ values, as can be checked numerically.
In order to avoid this artifact, we select one of the six faces of the sample,
and only pick the clusters contacting this face.
Moreover, we restrain the angular averaging process to $\textbf{Q}$ vectors parallel to the chosen surface. 
As expected, using this procedure suppresses the divergence at small $Q$ (except at $p_c$), at the cost of a larger statistical noise. The obtained structure factor corresponds to the typical experimental situation of Ref.\citenum{Page1995a}, where the incident wave vector is normal to the surface of the sample and the scattered radiation is observed at small angles close to the forward direction.

\begin{figure}[ht]
  \center
 \includegraphics[scale=0.3]{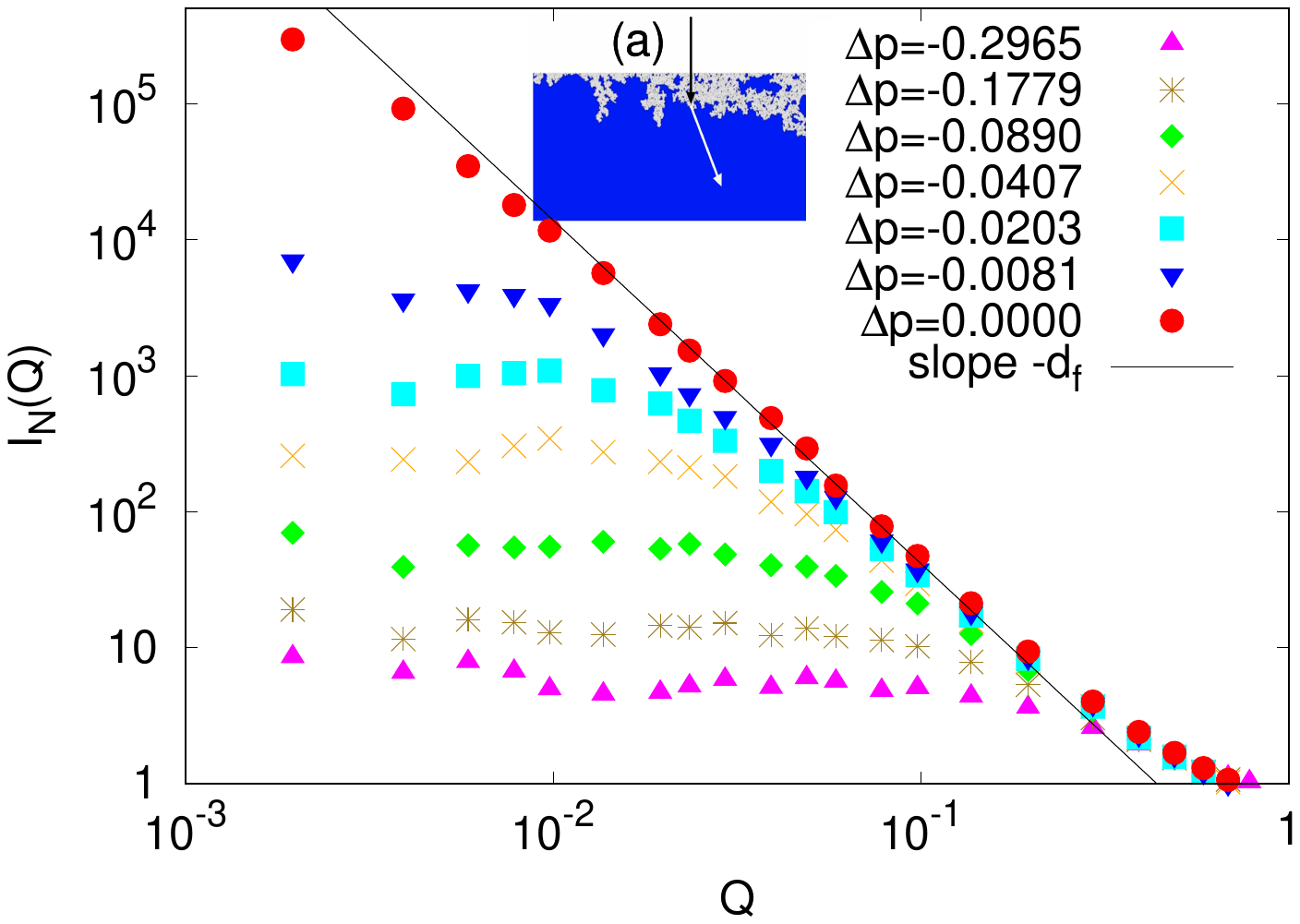}
\includegraphics[scale=0.3]{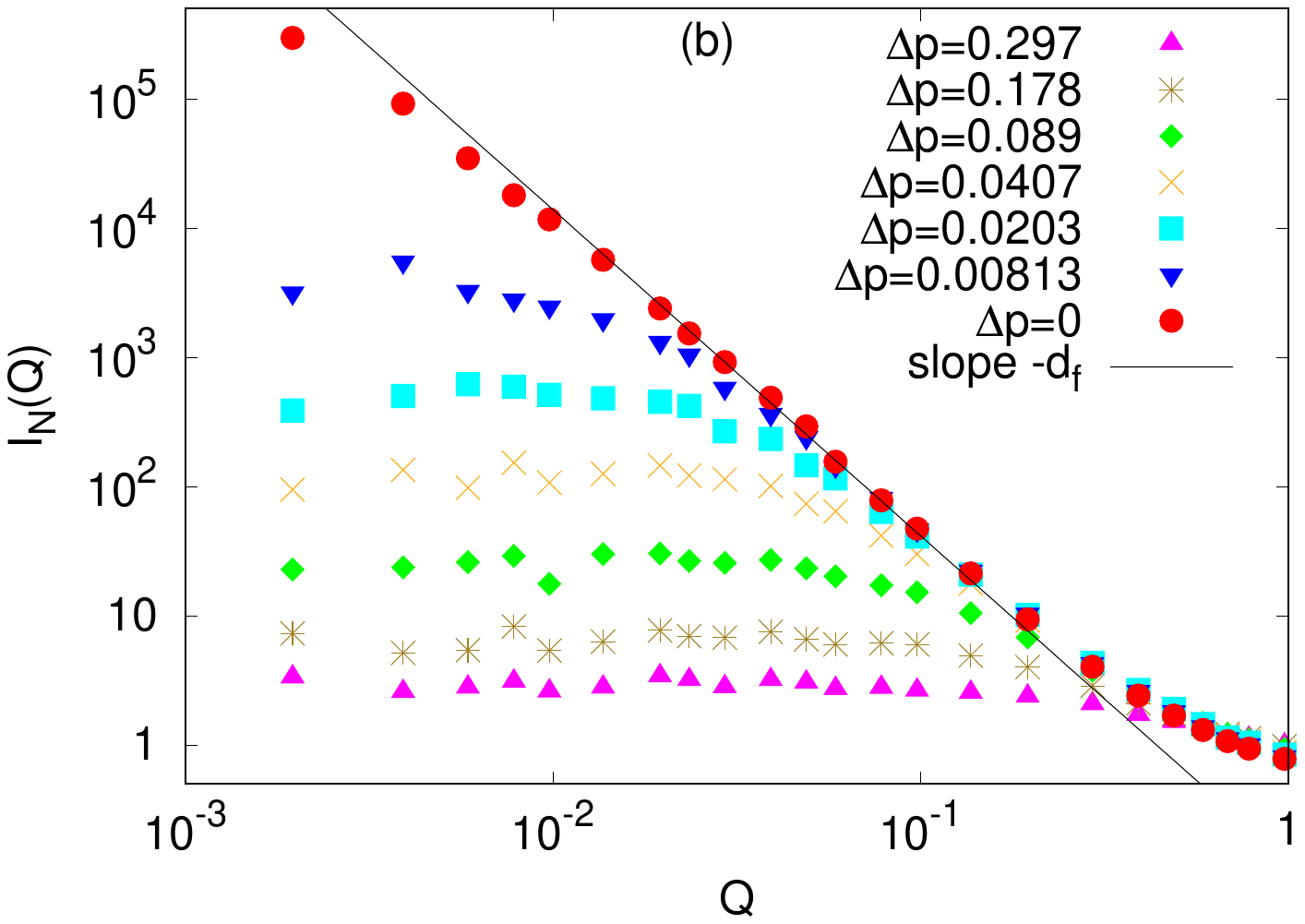}
  \caption{Normalized structure factor for invasion from a surface below (a) and above (b) $p_c$.  In (a), the inset shows the scattering geometry considered, where the incident wave vector is perpendicular to the surface. The data are noisier than in the germs case, due to the fact that the average is performed over a limited number of orientations of the $Q$ vector (see text). At percolation, a fractal behavior is obtained over a larger $Q$ range than in the germs case at fraction $10^{-7}$ (Fig.~\ref{fig:SO-Q-germs-belowpc}).
  }
  \label{fig:SQNboundaryclusters}
\end{figure}

\medskip
Figures~\ref{fig:SQNboundaryclusters}a and b show the resulting $Q$-dependence of
 the normalized structure factor below and above the percolation threshold, computed by
  averaging over the same four samples as in the germs case. 
   As in the case of invasion from bulk germs (Figs.\ref{fig:SO-Q-germs-belowpc}a and \ref{fig:SdeQaprespc}a), 
   the saturation value at low $Q$ values grows as percolation
     is approached on both sides of $p_c$.
      At $p_c$, $I_N(Q)$ varies close to $Q^{-d_f}$, as is the case for the larger cluster, or for a dilute distribution of bulk germs. 
    This behavior results from the fact that $I_N(Q)$ probes the
      whole volume of the sample: since clusters extend perpendicular to the surface
       only over a depth of order their radius of gyration, only the larger ones 
        contribute to $I_N(Q)$ far enough of the surface, explaining why, in contrast to bulk invasion, the fractal structure at $p_c$ is retained at large scale. At 2D, Fig.~\ref{fig:pictureboundaryclusters}  provides a pictural demonstration of this property. Except close to the invasion surface, the active sites at $p_c$ are those of the largest (percolating) cluster.
  This property explains why, as seen in Fig.~\ref{fig:Experiments}b \cite{Page1993a,Page1995a}, 
  the structure factor measured during evaporation from the surface of Vycor  does exhibit 
a  $Q^{-d_f}$ behavior, despite the large density of surface germs.

  \begin{figure}[ht]
 \includegraphics[scale=0.3]{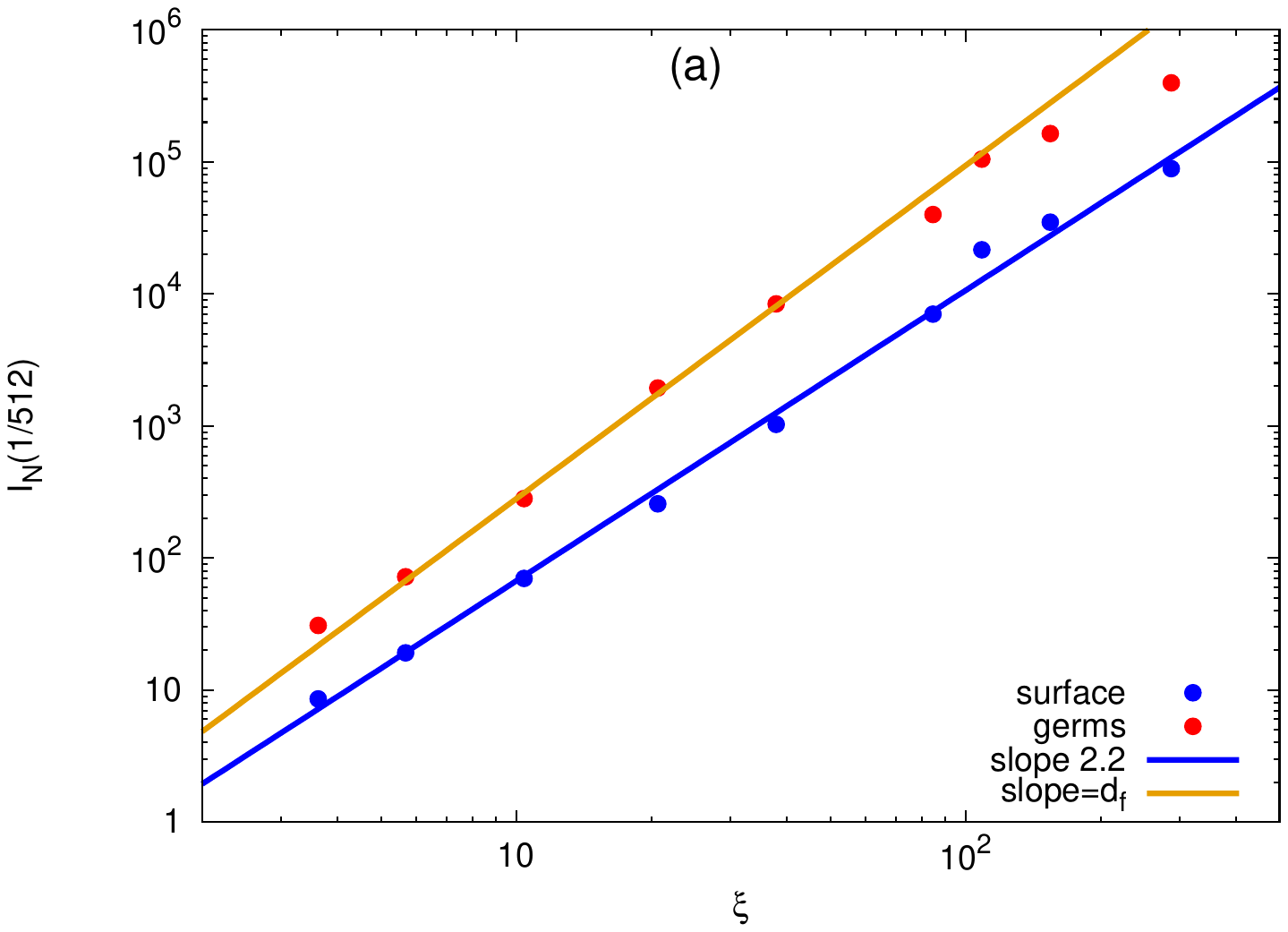}
\includegraphics[scale=0.3]{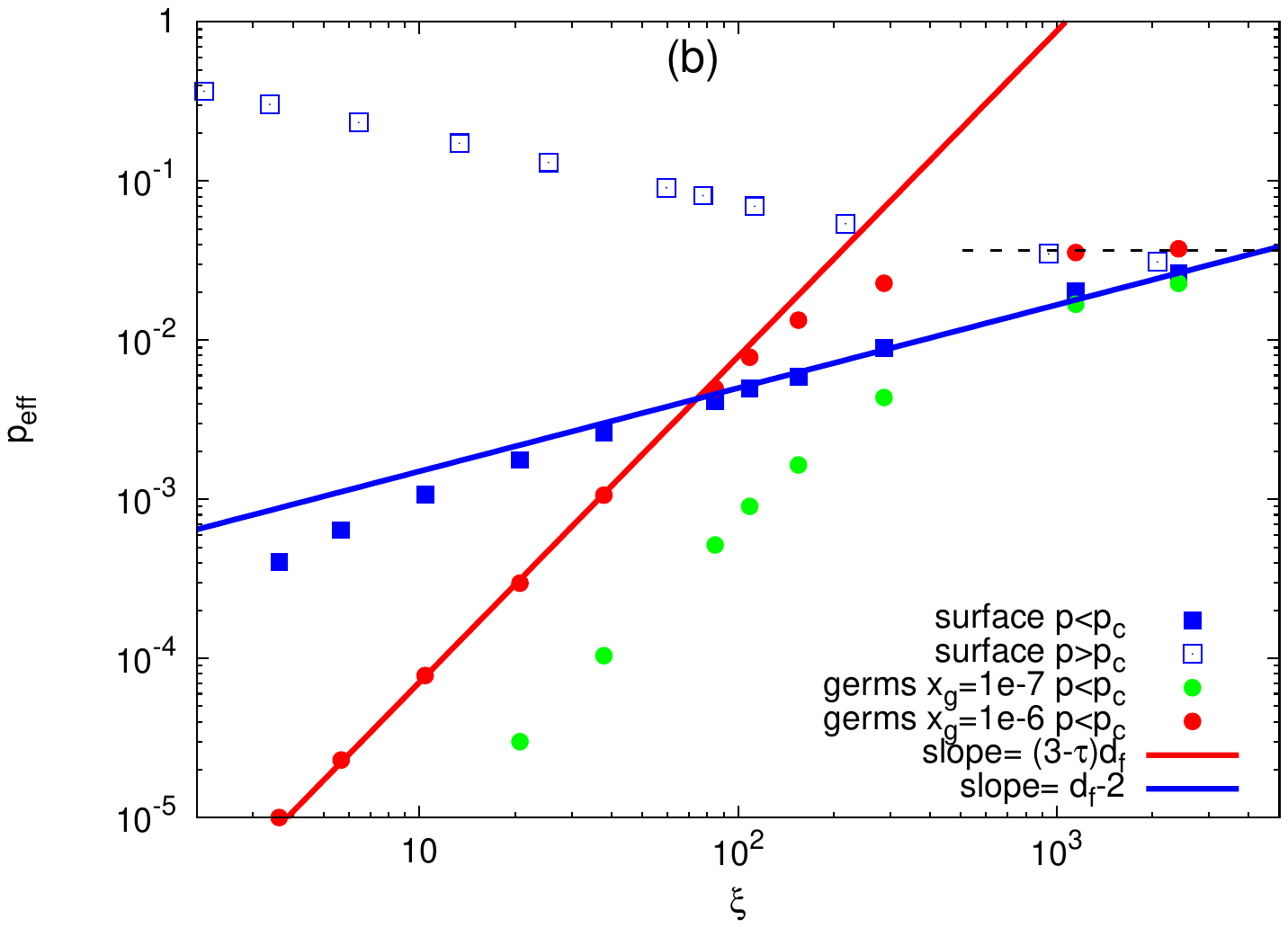}
  \caption{Comparison of invasion from bulk germs and surface; (a):  below $p_c$, $I_N(Q \to 0$) grows slower with $\xi$ for invasion from surfaces than from bulk germs  at $x_g=10^{-7}$; (b): $p_{\rm{eff}}$ for surface invasion below (filled symbols) and above (open symbols) $p_c$. Below $p_c$, $p_{\rm{eff}}$ grows approximately as  $\xi^{0.5}$, much slower than the $\xi^{3-2d_f}$ behavior observed for dilute bulk germs at $x_g=10^{-7}$ and $10^{-6}$.  At percolation, $p_{\rm{eff}} \approx L^{d_f-3}$, indicated by the horizontal dotted line, showing that the density is controlled by that of the percolating cluster.}
  \label{fig:SQNboundaryclusterslowQ}
\end{figure}

   \medskip
The large number of clusters of small size close to the surface has however consequences. 
  As shown by Fig.~\ref{fig:SQNboundaryclusterslowQ}a, away from $p_c$,
  the low $Q$ limit of $I_N(Q)$ for invasion from surfaces is smaller than
   for invasion from dilute germs.
    Moreover, Fig.~\ref{fig:SQNboundaryclusterslowQ}a also shows that $I_N(Q \to 0)$ increases with $\xi$ approximately as $\xi^{2.2}$, significantly slower than the $\xi^{d_f}$ behavior observed for germs in the dilute limit. 

  \medskip
It is interesting to compare the behavior of the absolute structure factor to that found for bulk invasion. 
Figure~\ref{fig:SQNboundaryclusterslowQ}b shows that,  below $p_c$,  
the $\xi$ dependence of $p_{\rm{eff}}$ is much slower than the $\xi^{3-2d_f}$ behavior found for bulk germs.
 For $\xi >30$, $p_{\rm{eff}} \sim \xi^{0.5}$. 
 This exponent is consistent with the simple assumption of a fractal structure up to $\xi$, confined in a layer of thickness $\xi$, which gives  $p_{\rm{eff}}~ \sim \xi^{d_f-2}/L$.
 This different behavior is responsible for the $p$ dependence of  $I(Q,p)$ below $p_c$ being different for the two situations of invasion, as seen in Fig.~\ref{fig:SQnorm}a and Fig.~\ref{fig:SQnorm}d.
However, Figs.~\ref{fig:SQnorm}a and \ref{fig:SQnorm}b, and   Figs.~\ref{fig:SQNboundaryclusterslowQ}b respectively show that, at $p_c$, both the normalized structure factor $I_N(Q)$ and $p_{\rm{eff}}$ are nearly identical for invasion from the surface, and for invasion from bulk germs at a very small fraction.
   As a result,  the absolute structure factor at $p_c$ is identical as well. 
   For our finite size sample, this identity results from the fact that, in both cases, all physical quantities are dominated by the largest cluster. 
   We expect that this would remain true for any size $L$, provided that, in the case of bulk germs, the germ fraction is low enough. For invasion from surfaces, the percolation threshold is also correctly determined by the maximum of the absolute structure factor, despite the locally large density of surface germs.

\medskip
We conclude that, at $p_c$, both the $Q$ dependence and the absolute value of the scattering signal should be similar for invasion from surfaces or dilute bulk germs. Differentiating the two processes thus requires to perform spatially resolved, local, measurements of $I_N(Q \to 0)$ as a function of the distance to the sample's surface.

\section{Conclusions}
\label{sec:conclusions}
Our systematic numerical study and its analysis allowed us to decipher the scattering properties of percolation clusters.
For single clusters, we demonstrated that their well-known fractal structure at the percolation threshold $p_c$ remains essentially unaffected in a wide range around the transition. To the best of our knowledge, this property had not been previously reported. For clusters assemblies, we answered the different questions raised in the introduction, 
enabling us to establish the scattering signatures of invasion percolation.

 \medskip
    First, for  invasion both from bulk germs and surfaces, we show that the structure factor at large wavevectors $Q$ decays as $Q^{-d_f}$, where $d_f$ is the fractal dimension associated with individual clusters.
   This result contrasts with that previously reported for clusters generated by gelation followed by dilution, a difference which we could trace to the different  cluster size distribution.
    For invasion from bulk germs, larger clusters are over-sampled with respect to the Bernoulli distribution relevant for dilution. The structure factor is then dominated by the largest clusters, giving a $Q^{-d_f}$ decay. Similarly, for invasion from surfaces, only the largest clusters exist far from the surfaces,
     leading to the same $Q^{-d_f}$ decay. In the latter case, our simulations show that, at percolation, this behavior extends down to $Q$=0. 
   Retrospectively, this confirms the interpretation of the earlier experiment by Page \textit{et al} \cite{Page1995a}, which ascribed to a percolation phenomenon the observed power law decay of the structure factor.
    
\medskip
Second, we show that the low $Q$ limit of the structure factor (once normalized by the structure factor for a random distribution of sites with the same density) behaves differently for the two modes of invasion. For bulk germs, as long as the clusters are well separated, corresponding to $\xi$ smaller than  the distance $d_g$ between germs, this limit varies  as $\xi^{d_f}$ as in the case of the largest cluster only.  In this regime, structure factor measurements may give access to  the critical exponent $\nu$ describing the divergence of $\xi$ at percolation. In contrast, for invasion from surfaces, we find that the divergence of the low $Q$ structure factor is slower, reflecting that clusters connected to the surfaces have a wider distribution of sizes than for bulk germs invasion. Barring this difference, the scattering signal peaks at percolation, with the same  $Q^{-d_f}$ dependence, both for invasion from surfaces and invasion from very dilute germs. Hence, experimentally distinguishing the two processes requires spatially resolved measurements.

\medskip
For invasion from bulk germs, we also discussed the effect of inter-cluster correlations close to the percolation threshold and beyond.  On both sides of the transition, for $\xi > d_g$, the structure factor and the density of active sites are found consistent with a blob picture, where the sample is densely filled by fractal blobs of typical extent $d_g$. For very large germs concentrations, intercorrelations nearly suppress the scattering enhancement at percolation.
For $p > p_c$, $\xi < d_g$, and not too small $d_g$,  the structure factor and the density of active sites are controlled by those of the percolating cluster. We presented approximate expressions parametrizing the structure factor as a function of  $d_g$,  $\xi$, and $Q$.
Based on these expressions, we could reproduce the results of earlier simulations of evaporation in a porous glass in a complex case where the density of bulk germs and the distance to percolation vary simultaneously,  demonstrating their usefulness.

\medskip
Finally, we note that , while our simulations were performed on a cubic lattice, the emerging physical conclusions can be expected to be of wider validity.
Our  results thus provide a new basis for analyzing light or neutrons scattering experiments in the vicinity of a percolation transtion.

\section{Appendix A}

\begin{table}[t!]
\begin{center}
\begin{tabular}{ c c c c c c c c c  }
\hline 
\hline
$p$ & 0.21920 & 0.25616 & 0.28388 & 0.29892 & 0.30526 & 0.30907 & 0.30970 & 0.31032\tabularnewline
$\Delta p$ & -0.29653 & -0.17792 & -0.08896 & -0.04069 & -0.02035 & -0.00812 & -0.00610 & -0.00411\tabularnewline
$\xi$ &3.3  &5.5 &10.5 &21& 40 &83& 105 &155\tabularnewline
\hline
$p$ & 0.31097 & 0.31147 & 0.31154 & 0.31160 & 0.31166 & 0.31173 & 0.31223 & 0.31288\tabularnewline
$\Delta p$ & -0.00202 & -0.00042 & -0.00019 & 0.00000 & 0.00019 & 0.00042 & 0.00202 & 0.00411\tabularnewline
$\xi$  &303  &350 &351 &334 & 313 &323& 107& 96\tabularnewline
\hline
$p$ & 0.31350 & 0.31413 & 0.31794 & 0.32428 & 0.33932 & 0.36704 & 0.40400 & \tabularnewline
$\Delta p$ & 0.00610 & 0.00812 & 0.02035 &  0.04069 &  0.08896 &  0.17792 &  0.29653 & \tabularnewline
$\xi$ & 50& 43 &23 &11.5& 6.0& 3.2& 2\tabularnewline
\hline
\hline
\end{tabular}
\end{center}
\caption{ Values of $p$, $\Delta p=p/p_c-1$ used in this work, and corresponding results for the correlation length $\xi$.
}
\label{table:pValues}
\end{table}

\begin{figure}[ht]
\includegraphics[scale=0.4]{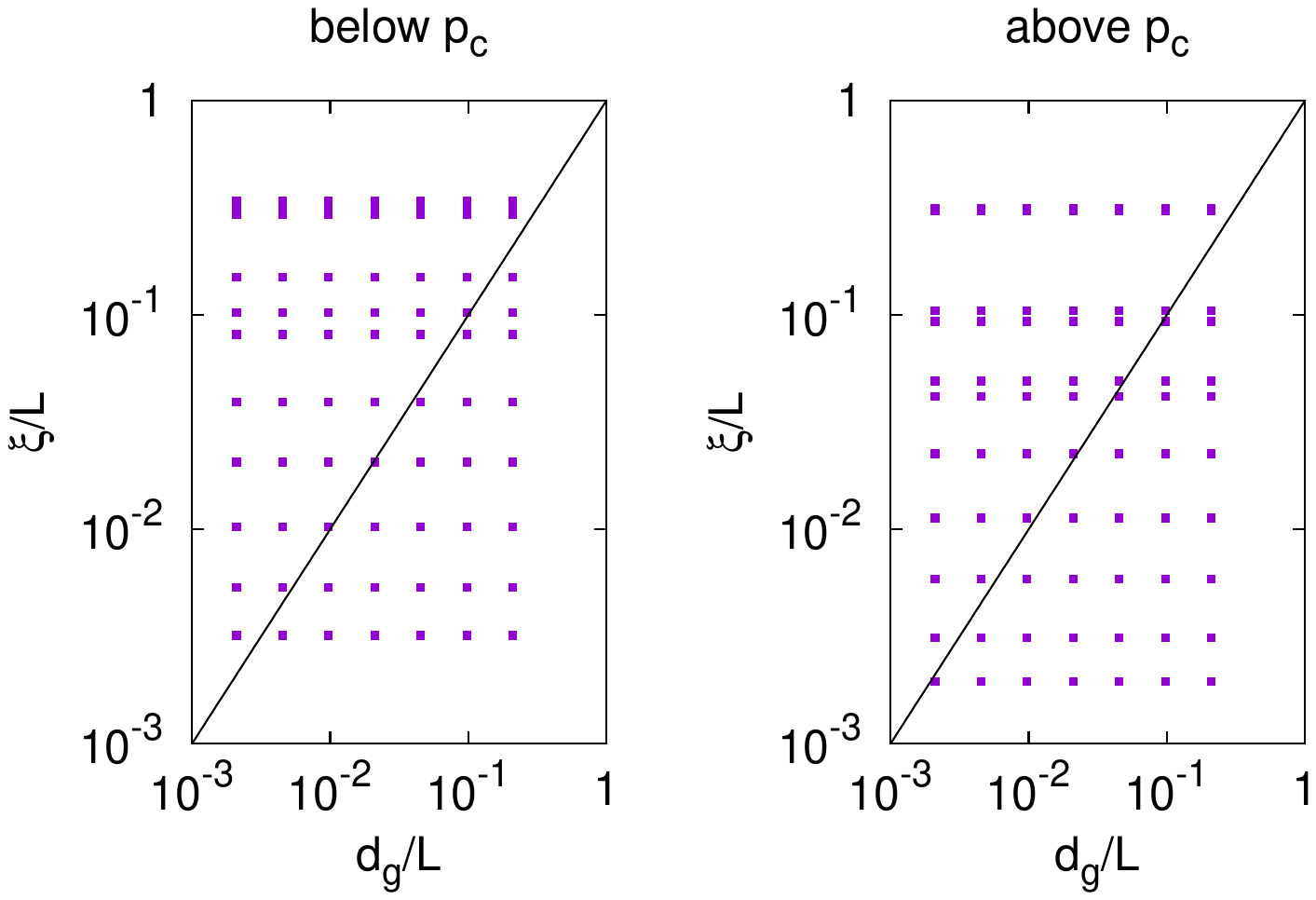}
\caption{Invasion from bulk germs: points of simulation in the plane ($\xi,d_g$) for which the structure factor has been computed. $\xi$ and $ d_g$ are given in units of $L$=1024. $d_g$ values correspond to germ fractions increasing from $10^{-7}$ up to  $10^{-1}$ by steps of 10. $\xi$ corresponds to the values of $p$ given in Table~\ref{table:pValues}. }
\label{fig:xidg}
\end{figure}

Simulations were performed for the 23 values of $p$ given in table~\ref{table:pValues}.
For each $p$, four configurations were generated and averaged. The averaged values of $\xi$ for the full cluster distribution, excluding the percolating one, are given in table~\ref{table:pValues}. When the normalization of the axes of figures involves $\xi$, these are the values used. 
Simulations for bulk germs were performed for seven germ fractions, ranging from 10$^{-7}$ to 10$^{-1}$ by steps of 10. 
Figure~\ref{fig:xidg} compares the corresponding average distance between germs, $d_g$, to  $\xi$.
 The straight line $\xi=d_g$ approximately separates the dilute and concentrated regimes.
  
\begin{figure}
    \includegraphics[scale=0.3]{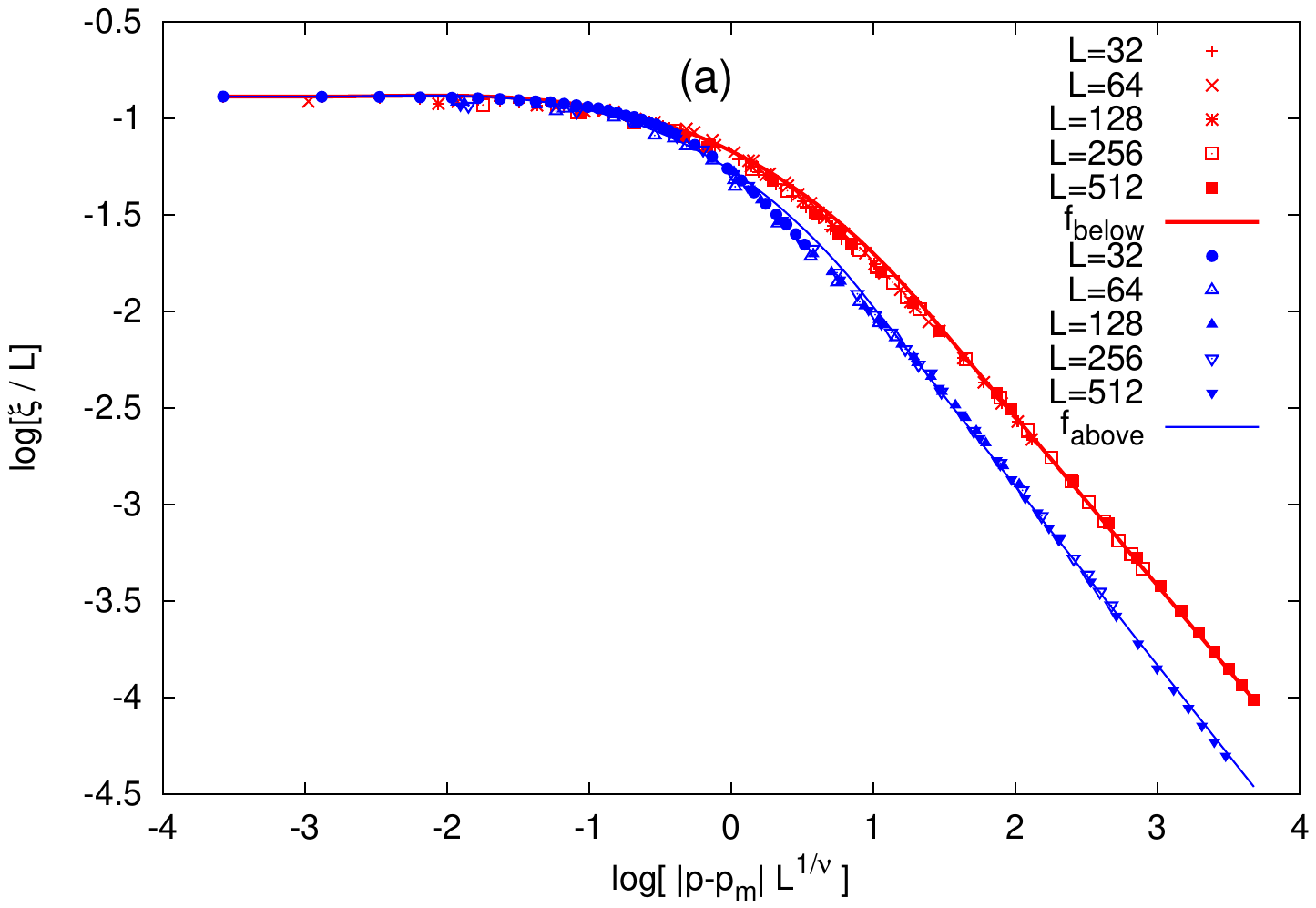} 
  \includegraphics[scale=0.3]{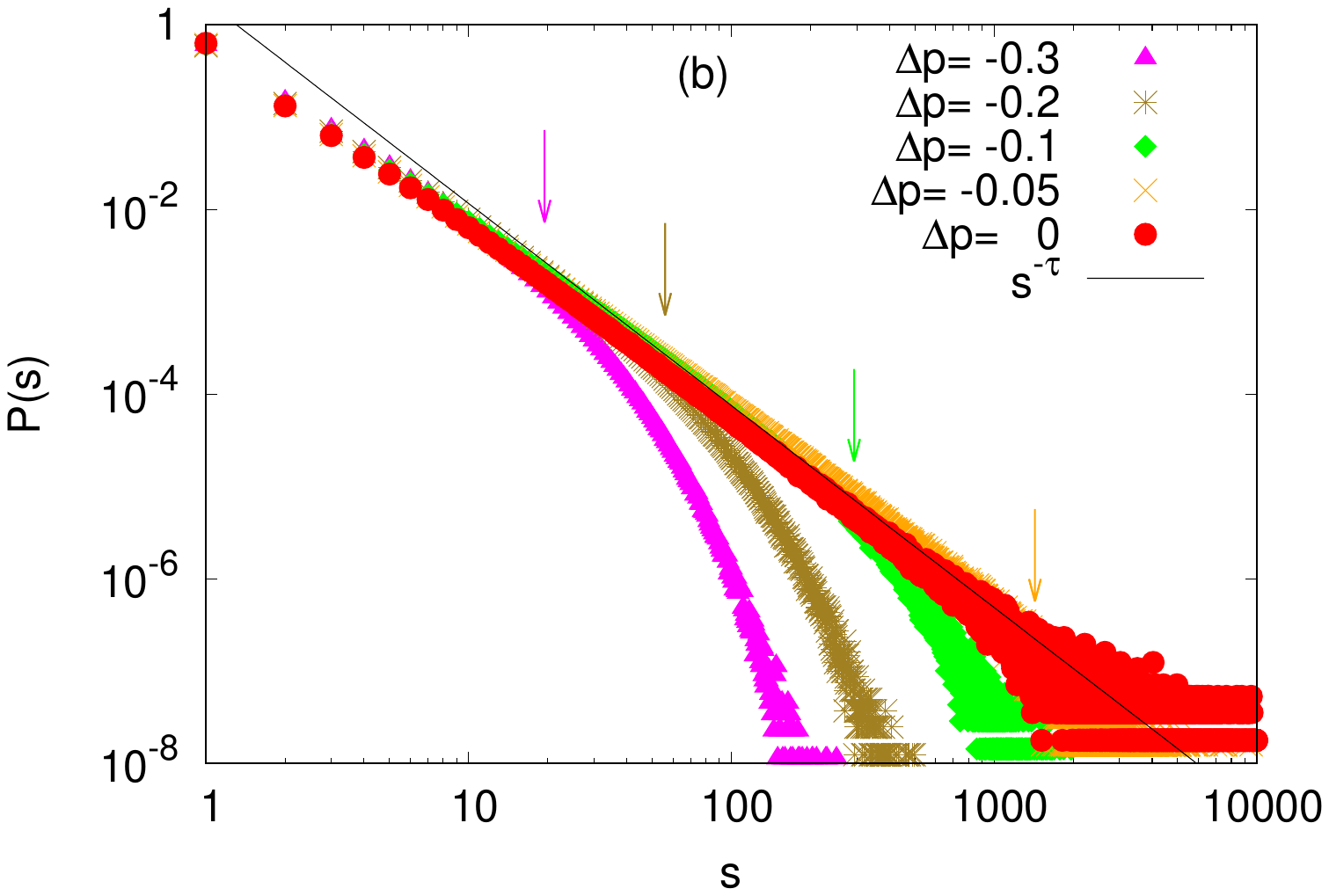}
 \caption{(a) Finite size scaling of the correlation length $\xi$ for all clusters except the percolating one. $p_m$ is the size dependent percolation threshold;(b) Cluster size distribution for different $p$ below $p_c$ for a 1024$^3$ cubic lattice. $\Delta p = (p-p_c)/p_c$. The black line corresponds to a power law with the theoretical exponent $\tau=1+d/d_f$=2.189. Arrows correspond to $s=\xi^{d_f}$, where $\xi$ is the correlation length for the full distribution of clusters.}
  \label{fig:xiPs}
\end{figure}

  \medskip
  Specific simulations using a large number of configurations were carried out to precisely study the  $p$ dependence of $\xi$ for lattices of linear size $L$ from 32 to 512.
Figure~\ref{fig:xiPs}a shows the results of these simulations, using standard finite size scaling coordinates.
For $\xi /L <  0.14 $, finite size effects are negligible and the behavior of $\xi$ in  a range of $\pm 30\%$ around $p_c$ 
 is well described by:
 
\begin{equation}
  \xi= \begin{cases}
    1.25 \,  (1-p/p_c)^{-0.876}   &   p<p_c \\
    0.68 \, (p/p_c-1)^{-0.929}  &   p>p_c
  \end{cases}
\end{equation}

where $\xi$ is in units of mesh size. These expressions are used in the paper to produce the continuous lines in Figs.\ref{fig:SQnorm}.  Note that a good fit requires the exponent of $|1-p/p_c|$  above $p_c$ to slightly differ from $\nu$, in contrast to the usual assupmtion  that the critical exponent is the same on both sides of the transition.  This might result from an off-critical effect due to the large range of $p$ values studied. 

\medskip
Finally, Fig.~\ref{fig:xiPs}b shows the cluster size distribution $P(s,p)$ deduced from our simulations for several values of $p$. In agreement with the litterature, $P(s,p_c)$ at percolation is accurately described by a power law distribution $s^{-\tau}$, with $\tau=1+d/d_f$. Below $p_c$, this behavior is cut-off at a size $s$ scaling approximately as $\xi^{d_f}$, consistent with the fact that the extent of the largest clusters scales as $\xi$.

\section{Appendix B}
\begin{figure}[ht]
  \includegraphics[scale=0.3]{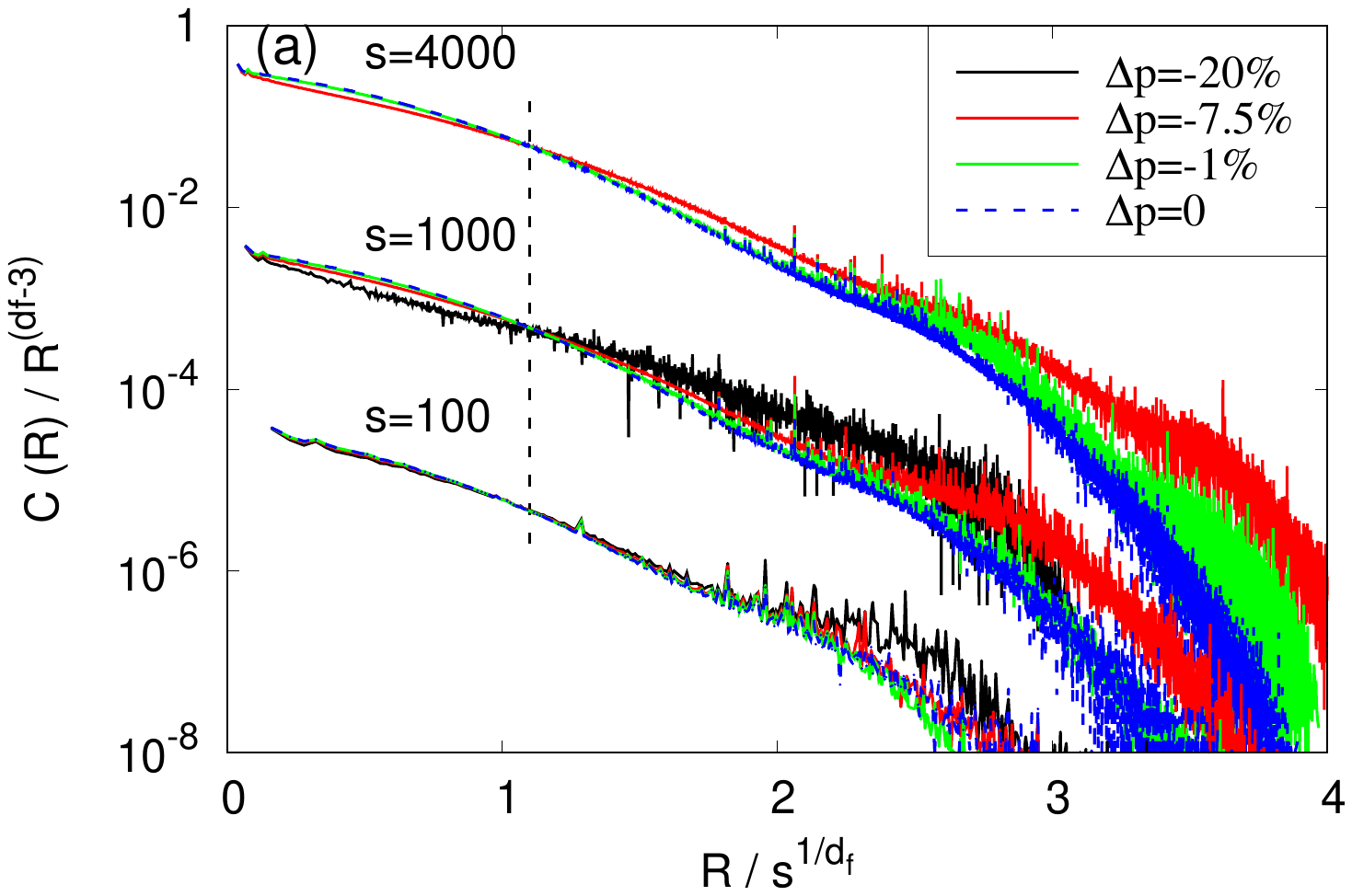}
 \includegraphics[scale=0.3]{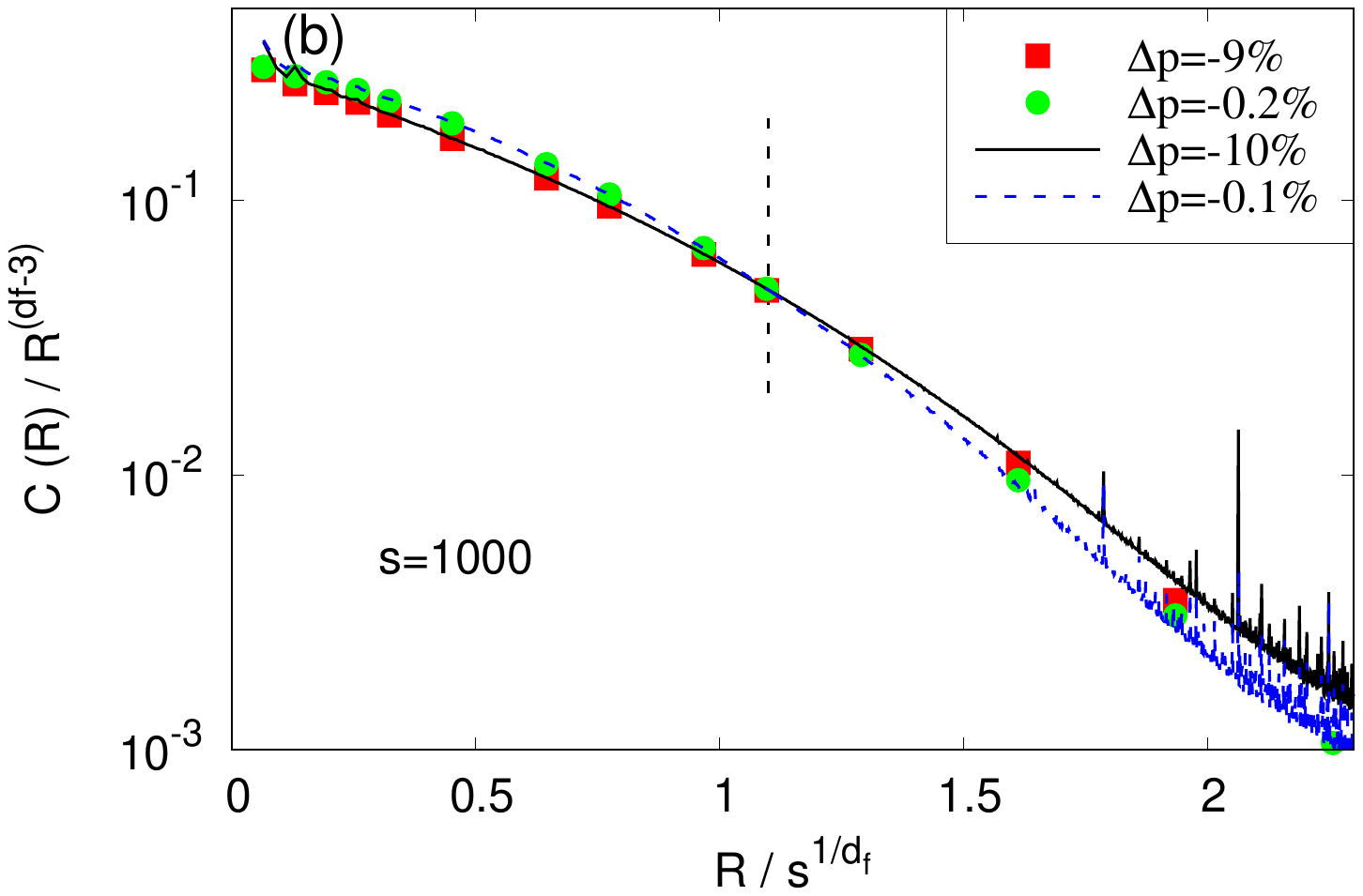}
 \caption{Correlation function for clusters of size $s$: (a) for clusters generated by an invasion algorithm with $s$=100, 1000, 4000, and different relative distances $\Delta p$ to the percolation threshold $p_c$. For clarity, the curves for $s$=1000 and 100 have been respectively scaled by 1/100 and 1/10000; (b)  Comparison of the $s=1000$ (within $\pm$ 2\%) case of Fig.~\ref{fig:correlInvasion}a to  $C_{1}(R)- p_{\rm{eff}}$ obtained for all Bernoulli clusters with same size $s$ (within $\pm$ 5\%) on the $L=1024$ lattice, at approximately similar values of $p$.  Lines correspond to the invasion algorithm and symbols to the result for Bernoulli clusters. In both figures, $C(R)$ is normalized by
 $R^{d_f-3}$, and $R$ by $s^{1/d_f}$, the typical extent of a fractal cluster of size $s$. The dashed lines show the position of the crossing point  $R^\star(s)$ discussed in the text.
 }
 \label{fig:correlInvasion}
\end{figure}

In this appendix, we directly study the  $p$ dependence of  the average correlation function $C(R)$ 
for a cluster of given size $s$ generated by an invasion process, and show that this dependence is modest.
 We further show that $C(R)$  for $s=1000$ coincides within a constant to the correlation function $C_{1} (R)$ obtained by a Fourier transform of the structure factor of the whole set of Bernoulli clusters of  size $s\pm \Delta s$ studied in \S\ref{sec:SizeSelection}.
This confirms that, as assumed in \S\ref{sec:SizeSelection}, the intercorrelation effects  for this set are negligible.

\medskip
To generate the clusters, we use a slightly modified version
of the invasion percolation algorithm \cite{Chandler1982}.
In this algorithm, the cluster is built site by site on an arbitrarily large cubic
lattice. To generate a cluster at $p$,
one first selects the origin and randomly chooses the parameter $p_i$ 
for the six neighbors of the origin. If all $p_i$ are larger
than $p$, the cluster is closed of size one. Otherwise, one adds to
the cluster the site with the lower $p_i$.  This creates new
neighboring sites $i$ for which one draws new parameters $p_i$.
One iterates until all neighboring sites have a parameter
larger than $p$ or the size exceeds the maximum size,
in which case the cluster is disregarded as not completed.
If the cluster is completed, it is kept only if it has the desired size.
Obviously, if $p$ is small, one has to generate a huge number of clusters to
obtain a large one. Conversely, if $p>p_c$,  most clusters will
not close, making it difficult to generate a cluster of intermediate size.

\medskip
Having obtained a collection of clusters of size $s$ with a small tolerance of $\pm$ 2\%, we obtain  the corresponding correlation $C(R)$ by counting for each cluster
the number of pairs of cluster sites separated by a distance $R$.
 $C(R)$ is
then averaged over all clusters of the collection, resulting in an excellent precision. 

\medskip
Figure~\ref{fig:correlInvasion} shows the correlation function $C(R)$ for clusters of
 different sizes ranging from 100 to 4000, and for different values
of $\Delta p$  ranging  from  0 to -0.2, the larger distance to $p_c$ for which 
 it was possible to generate clusters of the largest
size.
In this figure, the correlation function is normalized by its dependence at percolation,
 $C(R)=R^{d_f-3}$, and $R$ is normalized by $s^{1/d_f}$, the typical extent of a fractal cluster of size $s$.
 
 \medskip
In agreement with expectation, Fig.~\ref{fig:correlInvasion}a shows that, at percolation,  $C(R)/R^{d_f-3}$  behaves as a function of $R/s^{1/d_f}$. However, unlike often assumed, this function is not a simple single exponential, but is curved downwards. This explains our finding that $I_N(Q)$ for a single cluster (Fig.~\ref{fig:SO_scaled_1024_3D}a) does not coincide with the prediction of Ref.\citenum{Sinha1989a}. Away from percolation, $C(R)/R^{d_f-3}$ depends on $p$. Such a dependence is expected. Indeed, the Boltzmann weight of a cluster of size $s$ with $f$ frontier sites (the sites connected to the cluster, but not belonging to it) is proportional to $p^s. (1-p)^{f}$.  This implies that the cluster structure must depend on $p$. However,  Fig.~\ref{fig:correlInvasion}a shows that this dependence has no critical character at $p_c$ and remains modest in a large range around $p_c$. This extends to a wider range of $p$ and $s$ values the conclusion drawn in the main text from the study of $I_N(Q)$ for Bernoulli clusters of given size $s$ on a lattice of linear size $L=1024$. 

\medskip
An unexpected feature of  Fig.~\ref{fig:correlInvasion}a is the existence of a radius $R^\star(s)$ at which the correlation function is identical for all tested values of $p$. This crossing point, indicated by a dashed line on Fig.~\ref{fig:correlInvasion}a, accurately corresponds to $R^\star(s)\approx  s^{1/{d_f}}$  for $R\ge 1000$. It would be interesting to study whether this property could be rigorously demonstrated. 

\medskip
We finally compare the average correlation function $C(R)$ for a single cluster computed using the invasion algorithm to  $C_{1} (R)$,  obtained by an inverse Fourier transform of the structure factor for all Bernoulli clusters  of same size $s$. Since, at large distances, $C_{1} (R) \to p_{\rm{eff}}$,  Fig.~\ref{fig:correlInvasion}b more precisely compares  $C(R)$ and $C_{1} (R) - p_{\rm{eff}}$ for a size $s$=1000, and similar distances to $p_c$. Both calculations are in excellent agreement, including the existence and position of the crossing point. Since $C(R)$ and $C_{1} (R)$ only differ by an additive constant,  the normalized structure factor for $Q>0$ of the whole set of Bernoulli clusters of size $s$ coincides with the average normalized structure factor of a single cluster of same size. This validates the physical assumption made in section~\ref{sec:SizeSelection}.

\section{Appendix C}
In this appendix, we discuss the $Q$ dependence of $I(Q)$  at the percolation threshold for two different sets of clusters.
The first set includes a small representative fraction of all the Bernoulli clusters at $p_c$, corresponding to the assumption made by Martin and Ackerson \cite{Martin_PRA1985}, and the 
second includes the clusters selected by invasion from a very small fraction of bulk germs.
In both cases, we assume that the density of clusters is small enough for interference effects to cancel. 
The total structure factor is then computed by summing the structure factors  $I(Q,s)=s. I_N(Q,s)$ of the individual clusters of size $s$, weighted by the cluster size distribution function $N(s,p)$:

\begin{equation}
I_N(Q,p)=\frac{ \int_{1}^{\infty} N(s,p) I_N(Q,s)\,s\,ds}{\int_{1}^{\infty}\,N(s,p)\,s\,ds}
\label{eq:SQdiluteappendix}
\end{equation}

Figure~\ref{fig:SO_scaled_1024_3D} implies that, at given $Q$, $I_N(Q,s)$ behaves approximately as

\begin{equation}
I_N(Q,s)  = \begin{cases}
s           &\text{if $s<s_0(Q)$}\\
s_0(Q) &\text{if $s>s_0(Q)$}
\end{cases}
\end{equation}
with 
\begin{equation}
s_0(Q) \propto Q^{-d_f}
\end{equation}

\medskip
Similarly, in agreement with Fig.~\ref{fig:xiPs}b, we approximate the  Bernoulli distribution $P(s,p)$ by $s^{-\tau}$ with $\tau=1+d/d_f$  up to $s=s_{\rm max} (p) \simeq \xi^{d_f}(p)$ and zero above.
If the set of clusters is representative of the Bernoulli distribution, $N(s,p)=P(s,p)$, and

\begin{equation}\label{eq:integraleSinha_P2}
I_N\left(Q\right)=\begin{cases}
\frac{\tau -2}{3-\tau} \,s_{\rm max}^{3-\tau}              & s_{\rm max}<s_0\\
\frac{s_{0}^{3-\tau}}{(3-\tau)}-s_0 \,s_{\rm max}^{2-\tau}  & s_{\rm max}>s_0
\end{cases}
\end{equation}

\medskip
At percolation, $s_{\rm max} \to \infty$, and, because $\tau>2$, the first term of the R.H.S.  of
Eq.~\ref{eq:integraleSinha_P2} for $s_{\rm max}>s_0$ dominates the second one. $I_N(Q,p_c)$ then behaves at large $Q$ as
$s_{0}^{3-\tau} \propto Q^{-d_f (3-\tau)}$, i.e.  a power law with an exponent $\simeq 3-2d_f$, smaller in magnitude than the
fractal dimension $d_f\approx$ 2.52. 

\medskip
This result coincides with Martin and Ackerson one's, obtained by an exact
calculation in the case where the cut-off function in the two-points correlation function $g_s(r)$  is a gaussian.   
In fact, Martin and Ackerson's result directly reflects the spatial dependence of $C(R)$, the  intra-cluster correlation function averaged over the Bernoulli distribution, $C(R) \propto R^{2(d_{f}-d)}$\cite{Stauffer2018}.
Since Martin and Ackerson assume the structure factor to be given by the sum of the individual structure factors,  this structure factor coincides with the Fourier transform of $C(R)$, leading to a $Q$ dependence with an exponent $-d-2(d_{f}-d)= d-2 d_f$, equal to the exponent $d_f (\tau -3)$ above.

\medskip
However, this result does not hold for our case of invasion percolation, where $N(s,P)=s \,P(s,p)$.
 Due to the extra $s$ factor, the denominator in Eq.~\ref{eq:SQdiluteappendix}  is then controlled by the the $s_{\rm max}$ bound, yielding

\begin{equation}\label{eq:integraleSinha_Ps1}
I_N\left(Q\right)=\begin{cases}
\frac{3-\tau}{4-\tau} \,s_{\rm max}                                  & s_{\rm max}<s_0\\
\frac{3-\tau}{4-\tau}  \frac{s_{0}^{4-\tau}}{s_{\rm max}^{3-\tau}}+s_0  & s_{\rm max}>s_0
\end{cases}
\end{equation}

\medskip
At percolation, $ s_{\rm max}   \to \infty$, 
so that $I_N(Q)$ is given by
Eq.~\ref{eq:integraleSinha_Ps1} for $s_{\rm max}>s_0$. Because $\tau<3$, $I_N(Q) = s_0(Q) \propto Q^{-d_f} $. The fractal exponent is thus preserved by the
 averaging process.  
 This conclusion is  in agreement with our direct numerical simulations of $I_N(Q)$ (see Fig.~\ref{fig:SO-Q-germs-belowpc}a). 

 \medskip
Away from $p_c$, and in the  $Q \to 0$ limit,   $ s_{\rm max} $ is finite, and $s_0 \to \infty$, so that $I_N(Q)$ is given by
Eq.~\ref{eq:integraleSinha_Ps1} for $s_{\rm max}<s_0$.
Accordingly, in the dilute regime, the normalized structure factor $I_N(Q \to 0)$ 
   should diverge as $ \xi^{d_f}$. This is indeed in agreement with our simulations (Fig.~\ref{fig:SO-Q-germs-belowpc}a).

 \medskip
Note however than the quantity
    directly measured by a scattering experiment is the absolute structure factor
     $I(Q \to 0)= p_{\rm eff} (1-p_{\rm eff})I_N(Q \to 0)$. This quantity  involves the fraction of selected sites
      $p_{\rm eff}=  \int_{1}^{\infty} N(s,p) \,s\,ds$. Since $s N(s,p)$  decays as $s^{2-\tau}$,
       the latter integral diverges as  $\xi^{(3-\tau) d_f}$,
        also in agreement with our simulations (Fig.~\ref{fig:scalingpeff}). 
        As a consequence, in the dilute
     regime, the low $Q$ \textit{absolute} structure factor will diverge as  $\xi^{(4-\tau) d_f}$, approximately twice faster than the normalized structure factor. 
     Measurement of $d_f$ from the divergence of the low $Q$ absolute structure factor thus requires to take into account the extra-factor $(4-\tau)$=$3-3/d_f$.
     
%

\end{document}